\documentclass[prx,aps,twocolumn,superscriptaddress]{revtex4}

\usepackage{epsfig,wrapfig}
\usepackage{amssymb,amsfonts,amsmath}
\usepackage{afterpage}
\usepackage{color}

\newcommand{\s}{\sigma}

\newcommand{\beq}{\begin{equation}}
\newcommand{\eeq}{\end{equation}}
\newcommand{\bea}{\begin{eqnarray}}
\newcommand{\eea}{\end{eqnarray}}
\newcommand{\tab}{\hspace{5mm}}

\newcommand{\col}{\color{black}}
\newcommand{\cl}{\color{black}}

\begin{document}

\title{{\col Parsimonious} evolutionary scenario for the origin of allostery\\
 and coevolution patterns in proteins} 

\author{Olivier Rivoire}
\affiliation{Center for Interdisciplinary Research in Biology (CIRB), Coll\`ege de France, CNRS, INSERM, PSL Research University, Paris, France}

\begin{abstract}
\ \\

Proteins display generic properties that are challenging to explain by direct selection, notably allostery, the capacity to be regulated through long-range effects, and evolvability, the capacity to adapt to new selective pressures. An evolutionary scenario is proposed where proteins acquire these two features indirectly as a by-product of their selection for a more fundamental property, exquisite discrimination, the capacity to bind discriminatively very similar ligands. Achieving this task {\col is shown to} typically require proteins to undergo a conformational change. We argue that physical and evolutionary constraints impel this change to be controlled by a group of sites extending from the binding site. Proteins {\col can} thus acquire a latent potential for allosteric regulation and evolutionary adaptation {\col because of long-range effects that initially arise as evolutionary spandrels}. This scenario accounts for the groups of conserved and coevolving residues observed in multiple sequence alignments. However, we propose that most pairs of coevolving and contacting residues inferred from such alignments have a different origin, related to thermal stability. A physical model is presented that illustrates this evolutionary scenario and its implications. {\col The scenario can be implemented in experiments of protein evolution to directly test its predictions.}
\end{abstract}

\maketitle 

% keywords: Molecular discrimination| Allostery | Evolvability | Coevolution

{\col Proteins combine a capacity to perform exquisite tasks such as specific binding and catalysis with a capacity to adapt, both on physiological time scales through allostery and on evolutionary time scales through mutations. What principles underly this duality? How does it originate in the course of evolution? For proteins folding into a stable structure, a few generic properties constrain the explanations that we may provide. One is the ubiquity of long-range effects experimentally observed in all studied proteins: perturbations at a distance from the active site, whether in the form of mutations or of binding to another molecule, can affect significantly protein function~\cite{licata1995long,Daugherty:2000gz,Morley:2005kc}. Another is the coexistence of two types of patterns of coevolution revealed by statistical analyses of alignments of protein sequences: structurally connected groups of conserved and coevolving residues called sectors~\cite{Lockless:1999uf,Halabi:2009jc}, and structurally contacting pairs of residues distributed across the structure~\cite{Morcos:2011jg}. Finally, another feature reported to be essential to protein machineries is their capacity to undergo conformational changes~\cite{fersht1985enzyme}.}

{\col Here, we propose that a key to understand the link between high-performance and adaptability in proteins is to understand the physical and evolutionary implications of one of their most fundamental requirements: discriminative binding. Discriminative binding, the capacity to bind to a particular ligand but not to other similar ones,} is central to the function of many if not most proteins, from signaling proteins that respond to particular inputs and avoid cross-talk~\cite{pawson2000protein} to enzymes that bind to the transition state of a reaction but release its product~\cite{fersht1985enzyme}, transcription factors that recognize specific promoters among a profusion of similar motifs~\cite{todeschini2014transcription}, or antibodies that are highly specific to particular antigens~\cite{VanRegenmortel:1998uv}.

As binding takes place at a particular location on a protein surface, it is not {\it a priori} expected to be sensitive to distant perturbations, whether in the form of mutations or interactions with other molecules. But achieving exquisite discrimination  imposes particular constraints. Physically, we shall show how it typically involves a conformation change, or even a switch between two states that pre-exist any interaction with a ligand. Evolutionarily, {\col only few sequences can achieve discrimination which is shown to require a finely tuned binding site. This tuning} is generally difficult to accomplish based only on the few amino acids directly  interacting with the ligand. Additional tuning knobs are effectively provided by sites coupled to the binding site. Since only few residues interact directly with it, this generally implies recruiting distant sites. In this scenario, binding specificity is thus controlled through a conformational change by a group of sites extending beyond the binding site. This sensitivity of a functional phenotype to multiple and possibly distant sites endow proteins with a latent capacity to evolve allosteric regulation and adapt to new selective pressures. When integrating an additional constraint, thermal stability, we shall show that it also accounts for the different patterns of intra-protein coevolution that multiple sequence alignments of natural proteins report.

{\col We demonstrate this evolutionary scenario by means of simple physical models. These models are not meant to describe any particular protein but to capture the main constraints relevant to a discussion of long-range effects within stable protein folds: the fact that physical interactions are short-range. We analyze a variety of such physical models to ensure that our conclusions do not depend on the nature of the interactions: in all cases, we find that the physical implications of exquisite discrimination is a finely-tuned conformational switch between the ligand-free and the ligand-bound states. To examine the additional implications of evolutionary constraints, we then focus on one of the simplest models, a spin model where each site can be in just two states. This model can for instance be interpreted as describing side-chain fluctuations between a few discrete rotameric states in the context of a rigid backbone, one of the basic mechanisms by which long-range effects arise in proteins~\cite{dubay2011long}. We show that the evolutionary constraints cause with high probability a large part of the system to be involved in the discrimination. Sensitivity to perturbations distant from the active site thus arise as a by-product of a selection for discrimination, i.e., as an evolutionary spandrel~\cite{spandrel}. This side effect is shown to favor adaptation to new selective pressures when considering perturbations caused by mutations, and to enable the evolution of allostery when considering perturbations caused by interactions with other molecules. It is also shown to cause a generic coevolution pattern found in protein sequences, the extended groups of conserved and coevolving amino acids called sectors~\cite{Halabi:2009jc}. Another coevolution pattern is also generically found in protein sequences, which consists in isolated pairs of contacting amino acids distributed across the structure~\cite{Morcos:2011jg}. This pattern is shown to be captured by the model when introducing a constraint on thermal stability. The model thus provides an explanation for conformational switches, the origin of allostery, evolvability and the different patterns of coevolution observed in multiple sequence alignments as a consequence of only two selective constraints: discriminative binding and thermal stability. While it does not exclude other evolutionary scenarios that may lead to the same or different physical effects, it describes an interplay between evolutionary and physical constraints that is consistent with observations in many proteins and directly testable in experiments of protein evolution.}

\section{From discriminative binding to conformational changes}\label{sec:ex}

Formally, protein evolution involves three types of variables: (i) physical degrees of freedom $x$ describing the conformation of the molecule, (ii) evolutionary degrees of freedom $a$ defined by the protein sequence, and (iii) environmental degrees of freedom $\ell$ defined by the surrounding medium, here restricted to either the solvent or a ligand interacting at a particular binding site. These three variables are linked in a potential $U(x,a,\ell)$ that dictates through Boltzmann's law how the different physical conformations are sampled at thermal equilibrium: the probability of conformation $x$ is $P(x|a,\ell)= \exp[-\beta (U(x,a,\ell)- F(a,\ell))]$, where $F(a,\ell)=-\beta^{-1}\ln\int dx\exp [-\beta U(x,a,\ell)]$ represents the free energy of the system and $\beta$ the inverse temperature. 

A problem of discrimination arises when two ligands $\ell_r$ and $\ell_w$ can potentially be substituted for the solvent $\ell_0$ at the binding site, but only $\ell_r$ is desirable. Finding a sequence $a$ that solves this problem generally involves minimizing $\Delta F_r(a)= F(a,\ell_r)-F(a,\ell_0)$, the binding free energy to the right ligand $\ell_r$, while maximizing $\Delta F_w(a)= F(a,\ell_w)-F(a,\ell_0)$, the binding free energy to the wrong ligand $\ell_w$. When $\ell_r$ and $\ell_w$ are similar, this may lead to a trade-off. A precise formulation of this trade-off depends on the concentrations of the two ligands as well as on the relative cost and benefit that binding to them entail. {\cl Here, we consider a strong form of discrimination and} impose {\col $F(a,\ell_r)<F(a,\ell_0)<F(a,\ell_w)$. Under these conditions, binding with $\ell_r$ is favorable, $\Delta F_r(a)<0$, but binding with $\ell_w$ is unfavorable, $\Delta F_w(a)>0$. Different formulations of the same trade-off can also lead to the same conclusions (Appendix)}.

\begin{figure}[t]
\begin{center}
\includegraphics[width=\linewidth]{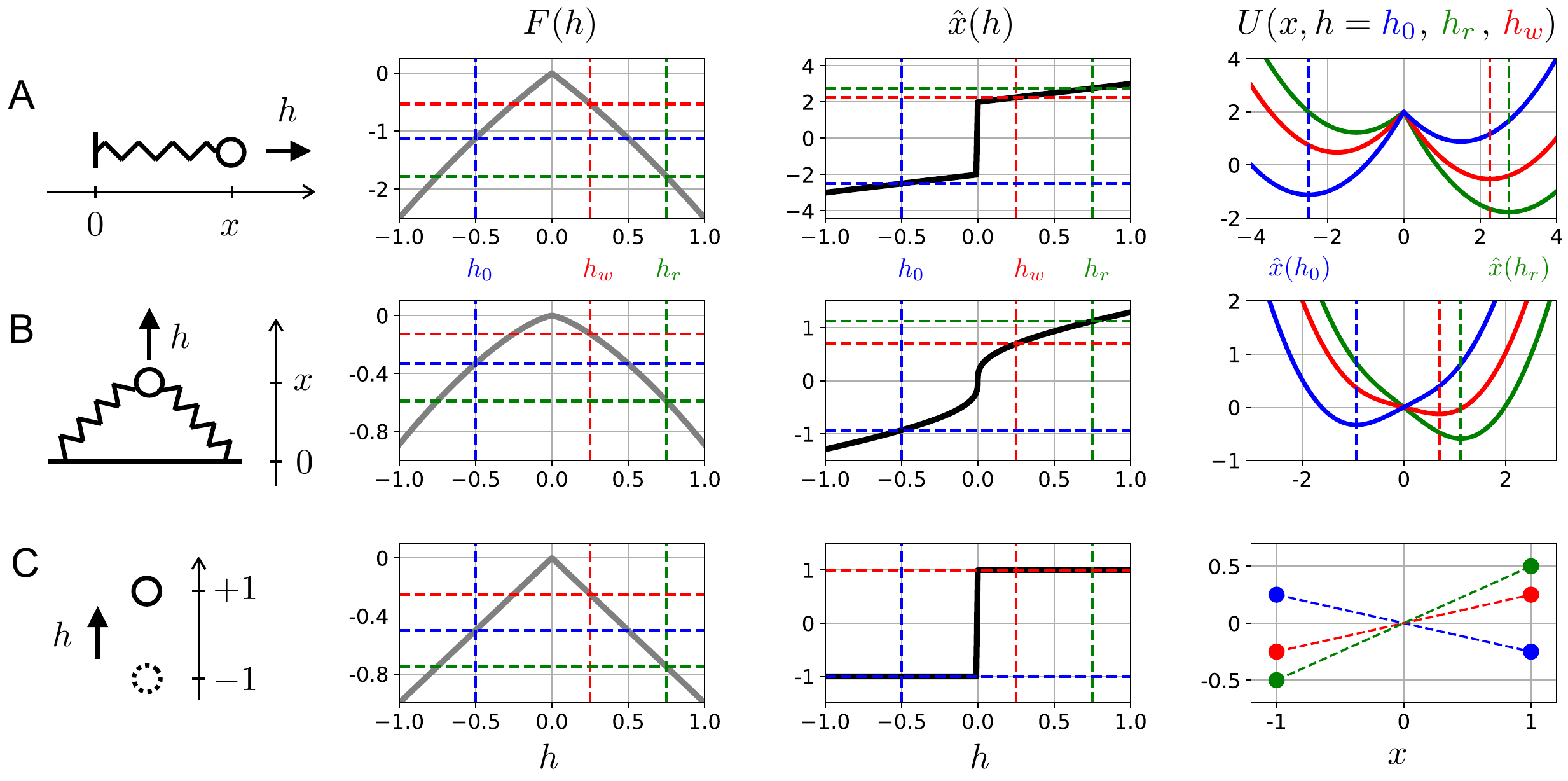}
\caption{Elementary models illustrating how exquisite discrimination implies evolutionary fine-tuning and a conformational change. {\bf A.} Model consisting of a single particle attached to a fixed point by a spring and subject to a constant force $h$. The potential is $U(x,h)=k(|x|-r)^2/2-hx$ and the free energy $F(h)$ is shown as a function of $h$ for $r=1$, $k=2$, $\beta=\infty$. The state $\hat x(h)$ of the system here is simply the minimum of $U(x,h)$. The constant force $h$ is jointly controlled by the evolutionary and environmental degrees of freedom $a$ and $\ell$ through $h=\ell-a$. To achieve exquisite discrimination in the form $F(a,\ell_r)<F(a,\ell_0)<F(a,\ell_w)$ the variable $a$ must be chosen so that $F(h_0=\ell_0-a)$ lies in between $F(h_w=\ell_w-a)$ and $F(h_r=\ell_r-a)$. When $\ell_0<\ell_w<\ell_r$, this requires $0\leq h_w<-h_0<h_r$. In this case, binding induces a conformational switch from $\hat x(h_0)<0$ to $\hat x(h_r)>0$. {\bf B.} Model with two springs under similar evolutionary constraints. Here, the conformational change is not a switch between two states of $U(x,h_0)$, although it is for other values of the parameters (Fig. S\ref{FigS:spring_var}). {\bf C.} Spin model where $U(x,h)=-hx$, exhibing the same phenomenology in a minimal setting where $x$ takes only two values $\pm 1$.\label{fig:elementary}}
\end{center} 
\end{figure}

Satisfying $F(a,\ell_r)<F(a,\ell_0)<F(a,\ell_w)$ when $\ell_r$ and $\ell_w$ are much more similar to each other than they are to $\ell_0$ is generally either impossible, or possible only for a restricted set of sequences $a$. This is best illustrated with a few elementary models. Consider first elastic networks, a mechanical framework commonly used as a coarse-grained description of proteins~\cite{Sanejouand:2013kj}. The simplest conceivable elastic network is a single mass attached to a fixed point by a spring. We assume here that it is additionally subject to a constant force  $h=\ell-a$ controlled by the evolutionary and environmental degrees of freedom $a$ and $\ell$ (Fig.~\ref{fig:elementary}A). The potential is therefore $U(x,a,\ell)=k(|x|-r)^2/2-(\ell-a)x$ where $x$ is the position of the mass along a dimension to which it is confined, $k$ the stiffness of the spring and $r$ its equilibrium length. In this model, $\ell$ and $a$ may take arbitrary real values. When $\ell_0<\ell_w<\ell_r$, it is readily shown that satisfying $F(a,\ell_r)<F(a,\ell_0)<F(a,\ell_w)$ requires $a$ to satisfy $(\ell_0+\ell_w)/2<a<(\ell_0+\ell_r)/2$ (Fig.~\ref{fig:elementary}A and Appendix). This corresponds to the mean conformation $\langle x\rangle_{a,\ell}=\int dx P(x|a,\ell)x$ switching from a negative to a positive value when the solvent $\ell_0$ is replaced by the ligand $\ell_r$. Importantly, this conformational switch is required only when $\ell_w$ and $\ell_r$ are sufficiently similar: if $\ell_w<\ell_0<\ell_r$, the solutions do not involve a change of sign (Appendix).

In this first model, achieving exquisite discrimination involves a switch between two states that are local minima of the potential $U(x,a,\ell_0)$ (Fig.~\ref{fig:elementary}A). This, however, need not be the case as seen by considering an harmonic potential $U(x,a,\ell)=k(x-r)^2/2-(\ell-a)x$, which also requires $a$ to verify $(\ell_0+\ell_w)/2<a<(\ell_0+\ell_r)/2$ but cannot sustain multiple states (Appendix). In this model, a conformational change nevertheless occurs upon binding with amplitude $\Delta x=|\ell_r-\ell_0|/k$. A model with the same phenomenology can also be constructed from two springs (Fig.~\ref{fig:elementary}B). Varying one parameter in this model, we can continuously interpolate between a one-state and a two-state model (Fig. S\ref{FigS:spring_var}). 

Exquisite discrimination can in principle be achieved without a conformational change, by a rigid lock-and-key mechanism, but fine-tuning the evolutionary parameters is in any case necessary (see shape space model in Appendix). A conformational change is, however, expected as soon as a minimal form of flexibility is present. A limiting case is a spin model where $x$ takes only two values, $x=\pm 1$. With $U(x,a,\ell)=(a-\ell)x$, the discrimination problem takes again the same form: when $\ell_0<\ell_w<\ell_r$, $a$ must be tuned to satisfy $(\ell_0+\ell_w)/2<a<(\ell_0+\ell_r)/2$ for the condition $F(a,\ell_r)<F(a,\ell_0)<F(a,\ell_w)$ to be fulfilled, in which case binding induces a conformational switch from $\langle x\rangle_{a,\ell_0}<0$ to $\langle x\rangle_{a,\ell_r} >0$ (Fig.~\ref{fig:elementary}C and Appendix). Graphically, the conformational change is again linked to the need for $\ell_0$ and $\ell_r$ to be associated with different ``branches'' of the free energy ($h<0$ and $h>0$ for $F(h=a-\ell)$ in Fig.~\ref{fig:elementary}).

In summary, achieving exquisite discrimination requires a flexible system to be evolutionarily tuned and to physically change conformation upon binding. In proteins, however, the evolutionary degrees of freedom are not controlled by continuous variables but by a limited set of twenty amino acids. At the binding site, relatively few values of $a$ are thus available to tune $F(a,\ell)$. As shown below, a generic solution is to enlarge the evolutionary space by coupling the binding site to other sites. Interestingly, having multiple tuning knobs not only favors adaptation to a particular selective constraint but also to alternative constraints. Besides, if distant sites are thus involved, they are likely to be allosteric: a perturbation at those sites, such as an interaction with another molecule, will alter the binding properties. These implications of evolutionary and physical constraints are independent of the underlying mechanisms {\cl and may be demonstrated for a wide class of potentials~\cite{Zadorin19}}. We therefore illustrate them using a physical model with the simplest form of flexibility, a spin model.

\section{Two-dimensional spin model}\label{sec:spin2d}

\begin{figure}[t]
\begin{center}
\includegraphics[width=\linewidth]{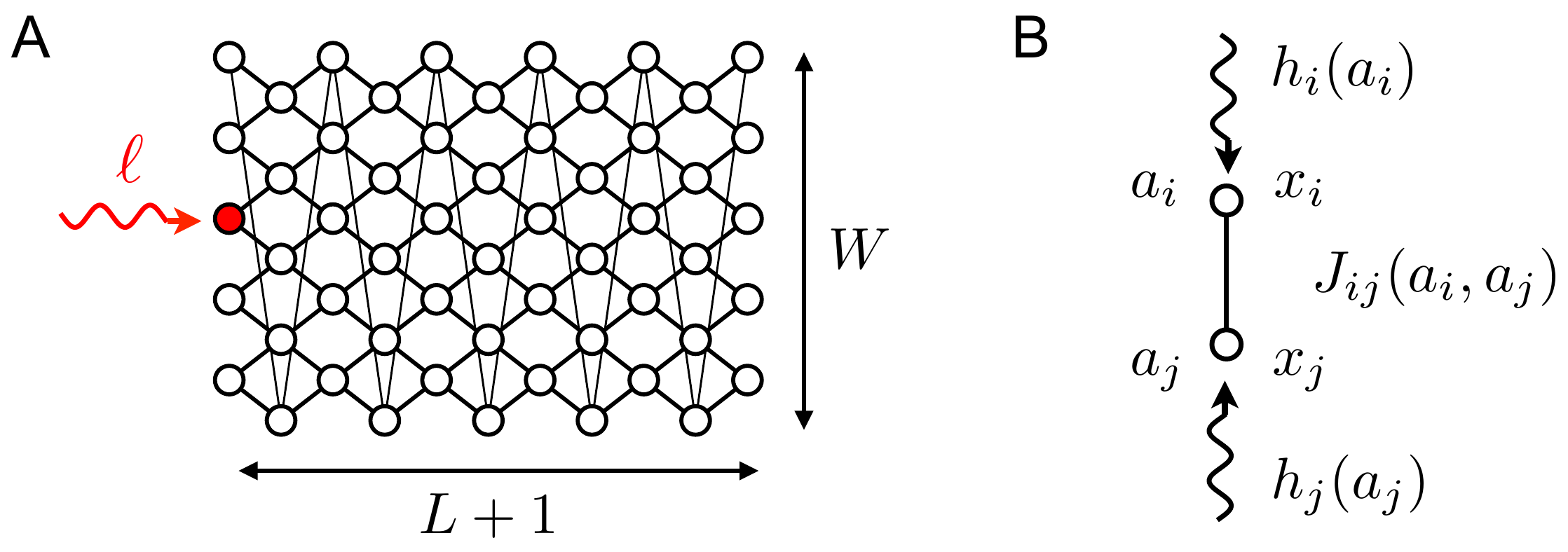}
\caption{Two-dimensional spin model. {\bf A.} Lattice on which the model is defined.  Each node $i$ carries a physical variable $x_i\in\{-1,+1\}$ and an evolutionary variable $a_i\in\{1,\dots,q\}$. One node at a boundary, here in red, defines the binding site, which may be subject to an external field $\ell$ representing a ligand. The thiner lines from top to bottom nodes implement the periodic boundary conditions, giving to the structure the geometry of a cylinder. Here and in the following figures, $L=10$ and $W=5$.  {\bf B.} The evolutionary variables $a_i$  and $a_j$ define fields $h_i(a_i),h_j(a_j)$ and couplings $J_{ij}(a_i,a_j)$ to which every physical variables $x_i$ and $x_j$ connected by a link in the lattice are subject. The potential $U(x,a,\ell)$ is given by Eq.~\eqref{eq:Uxal}. \label{fig:scheme}}
\end{center} 
\end{figure}

As a simple model with a non-trivial geometry, we consider a spin model defined on a two-dimensional lattice with periodic boundary conditions along one dimension (Fig.~\ref{fig:scheme}A). {\col This model is similar to the model introduced in Ref.~\cite{Hemery:2015ei} with two differences. First, for simplicity, we consider binary variables (spins) rather than continuous variables. Second and most importantly, the evolutionary scenario that we consider is totally different: instead of selecting explicitly for a long-range effect by varying the boundary conditions at the two extreme sides of the lattice, we select for binding specificity by varying the boundary conditions at a single site of the lattice. The point is indeed to show that long-range effects can evolve spontaneously with significant probability from a selection for exquisite discrimination, even though this selection is very localized.}

Each node $i$ is associated with two variables, a physical variable $x_i$ that can take two values $x_i=\pm 1$, and an evolutionary variable $a_i$ that can take $q$ values $a_i=1,\dots,q$. The evolutionary variables determine the fields $h_i(a_i)$ and couplings $J_{ij}(a_i,a_j)$ to which the physical variables are subject (Fig.~\ref{fig:scheme}B). Finally, a site $b$ is chosen to represent the binding site, which is subject to an additional external field $\ell$ representing an interaction with a ligand or with the solvent (red node in Fig.~\ref{fig:scheme}A). The potential is
\beq\label{eq:Uxal}
U(x,a,\ell)=-\sum_{\langle i,j\rangle}J_{ij}(a_i,a_j)x_ix_j-\sum_ih_i(a_i)x_i-\ell x_{b}
\eeq
where $x\in\{-1,+1\}^N$, $a\in\{1,\dots,q\}^N$ and $\ell\in\mathbb{R}$, $N$ being the total number of sites. {\col The couplings $J_{ij}(a_i,a_j)$ are possibly non-zero only between nearest neighbors in the two-dimensional lattice shown in Fig.~\ref{fig:scheme}A, so as to reflect the short-range nature of physical interactions.} The free energy given a sequence $a$ and a ligand $\ell$ is as usual $F(a,\ell)=-\beta^{-1}\ln\sum_x \exp(-\beta U(x,a,\ell))$ with $\beta$ the inverse temperature.

As a score of the extent to which a sequence $a$ discriminates the right ligand $\ell_r$ from the wrong ligand $\ell_w$, we consider
\beq\label{eq:phia}
\phi(a)=\min(-\Delta F_r(a),\Delta F_w(a)),
\eeq
where $\Delta F_r(a)=F(a,\ell_r)-F(a,\ell_0)$ and $\Delta F_w(a)=F(a,\ell_w)-F(a,\ell_0)$ are the binding free energies to the two possible ligands. With this formulation, $F(a,\ell_r)<F(a,\ell_0)<F(a,\ell_w)$ is equivalent to $\phi(a)>0$. {\col Other fitness functions implementing the same trade-off between $\Delta F_r(a)$ and $\Delta F_w(a)$ may also be used and lead to similar conclusions (Fig.~S\ref{FigS:fit}).} We sample sequences $a$ with probability $P(a)\propto e^{\gamma\phi(a)}$ using a Metropolis Monte Carlo algorithm~\cite{metropolis53}, which corresponds to an evolutionary dynamics in the origin-fixation limit, with $\phi(a)$ interpreted as a fitness function and $\gamma$ as an effective population size~\cite{McCandlish:2014us} (see Appendix for details).

For illustration, we consider a cylinder with $L=10$ layers of couplings and $W=5$ nodes per layer (Fig.~\ref{fig:elementary}A) for a total of $N=W(L+1)=55$ sites, in the range of smallest folding protein domains. With this geometry, the free energy $F(a,\ell)$ and other thermodynamical quantities can be computed exactly by transfer matrices~\cite{baxter82}. We take $q=5$, smaller than the number of natural amino acids but sufficient to generate a large evolutionary space of size $q^N\simeq 10^{38}$. To define $h_i(a_i)$ and $J_{ij}(a_i,a_j)$, we draw their values at random from normal distributions, $h_i(a_i)\sim\mathcal{N}(0,\s_h^2)$ and $J_{ij}(a_i,a_j)\sim\mathcal{N}(0,\s_J^2)$, independently for each $i,j$ and each $a_i,a_j$. Fixing the energy scale by setting $\beta=1$, we take $\s_h=\s_J=3$ so that some fields and couplings may be large relative to $\beta^{-1}$, as in natural proteins where the strength of some physical interactions, e.g., covalent bounds and steric constraints, can significantly exceed the scale of thermal fluctuations. We take $\ell_0=0$, $\ell_w=2$ and $\ell_r=3$, a choice of parameters for which $\phi(a)>0$ is achievable with a single spin, provided the adequate evolutionary diversity is available. Lastly, we take $\gamma=100$ to sample near-optimal values of $\phi(a)$.

\begin{figure}[t]
\begin{center}
\includegraphics[width=\linewidth]{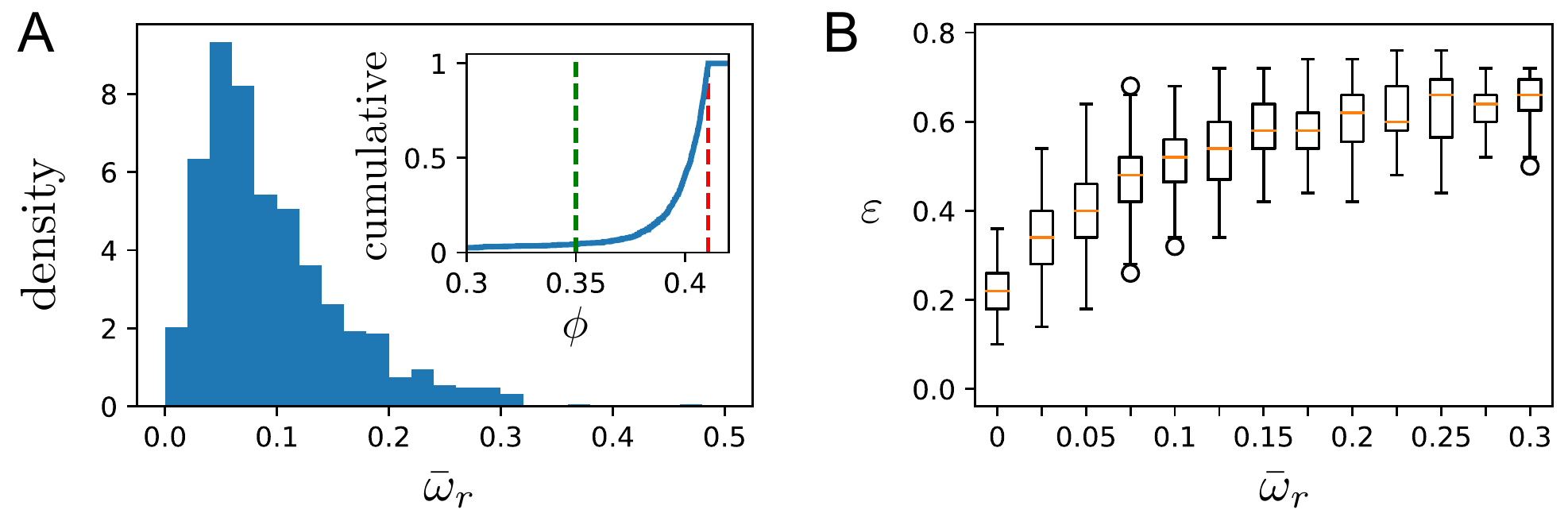}
\caption{Extent of the conformational change and evolvability for different realizations of the two-dimensional spin model. {\bf A.} Distribution of $\bar\omega_r$, the effective fraction of sites changing conformation upon binding, over 1000 different evolved systems $a$. Only the fittest systems with $\phi>0.35$ are considered. As the distribution of $\phi$ is picked around its theoretical maximum $\phi\simeq 0.41$, these top systems represent $>94\%$ of the total. For these systems, $\bar\omega_r$ is nearly independent of $\phi$ (Fig. S\ref{FigS:w_phi}A). Inset: cumulative distribution of $\phi$ over all 1000 systems. The red dotted line indicates the theoretical maximum. {\bf B.} For the same systems, the measure of evolvability $\varepsilon$ is positively correlated with $\bar\omega_r$, although independent of $\phi$ (Fig.~S\ref{FigS:w_phi}B): the more extended the conformational switch, the wider the range of phenotypes available in the neighborhood of a sequence, irrespectively of its fitness.\label{fig:Nheff}}
\end{center} 
\end{figure}

\begin{figure}[t]
\begin{center}
\includegraphics[width=\linewidth]{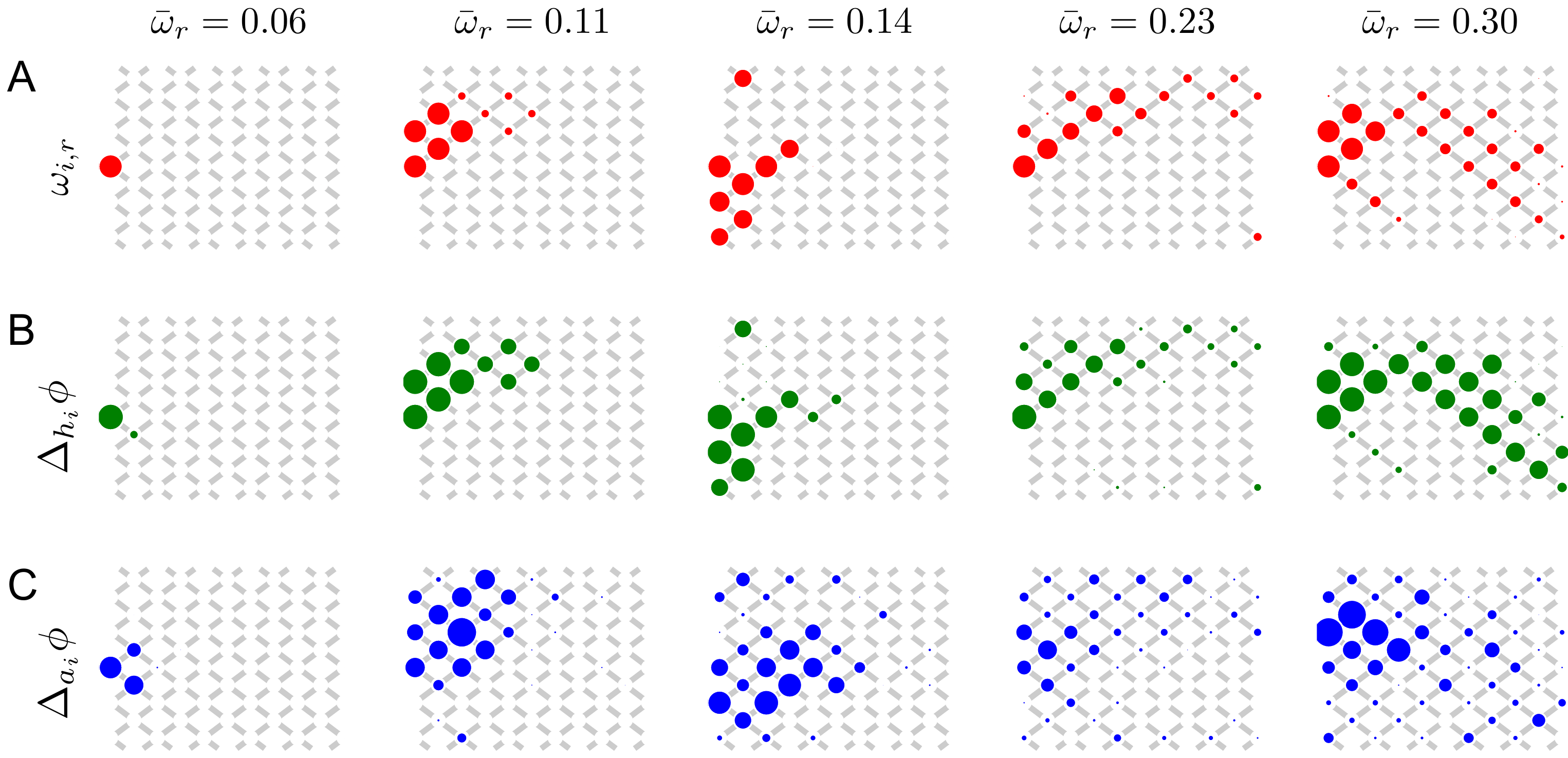}
\caption{Five evolved systems with equivalent fitness $\phi\simeq 0.4$ but switches of different extensions $\bar\omega_r$, as indicated on the top. {\bf A.} Conformational change $\omega_{i,r}$ at each position $i$ when $\ell_r$ is substituted for $\ell_0$. {\bf B.} Maximal fitness cost $\Delta_{h_i}\phi$ when adding an external field $h_i=\pm 2$. {\bf C.}  Average fitness cost $\Delta_{a_i}\phi$ of mutating $a_i$. In each case, the size of the dots is proportional to the quantity of interest. The first two rows are nearly identical, indicating that the same sites that change conformation are allosteric. The relation to sites displaying a strong mutational effect (last row) is also apparent although less straightforward due to the impact of mutations on both $h_i(a_i)$ and $J_{ij}(a_i,a_j)$. These examples illustrate how long range effects may arise from local selection for exquisite discrimination (the binding site is always the middle node on the left edge as in Fig.~\ref{fig:scheme}A).\label{fig:ex}}
\end{center} 
\end{figure}

With these parameters, most evolutionary trajectories lead to $\phi(a)>0$ after $1000$ iterations ($>98\%$, see inset of Fig.~\ref{fig:Nheff}A). To visualize and quantify conformational changes induced by substituting $\ell_r$ for $\ell_0$, we introduce 
\beq\label{eq:omega}
\omega_{i,r}(a)=\frac{1}{2}\left|\langle x_i\rangle_{a,\ell_r}-\langle x_i\rangle_{a,\ell_0}\right|,\quad\bar\omega_r(a)=\frac{1}{N}\sum_{i=1}^N\omega_{i,r}(a)
\eeq
where $\langle x_i\rangle_{a,\ell}=\int dx P(x|a,\ell)x_i$ stands for the mean value of $x_i$ given $a,\ell$. $\omega_{i,r}(a)$ quantifies the extent to which site $i$ undergoes a conformational change upon binding to the right ligand: $\omega_{i,r}(a)=0$ indicates no change, while $\omega_{i,r}(a)=1$ indicates a maximal change between two polarized states. The site-averaged quantity $\bar\omega_r(a)$ may be interpreted as an effective fraction of sites taking part in the switch ($\bar\omega_{w}(a)$ is similarly defined by considering $\ell_w$ instead of $\ell_r$). As shown in Fig.~\ref{fig:Nheff}A, $\bar\omega_r(a)$ varies widely from one evolved system to the next, even among systems with nearly identical fitness value $\phi(a)$. In most but not all cases, the set of sites involved in the switch extends beyond the binding site. The size and shape of this extension varies again from case to case (Fig.~\ref{fig:ex}). As in the simpler models, a switch evolves only under a constraint for exquisite discrimination controlled by the similarity between the two ligands $\ell_r$ and $\ell_w$: if $\ell_w<\ell_0<\ell_r$ no switch evolves while if $\ell_0<\ell_w<\ell_r$ or $\ell_0<\ell_r<\ell_w$ it always does (Fig.~S\ref{fig:var_hp}B).

Consistent with the proposed scenario, sites involved in the switch, and only those sites, are sensitive to perturbations. This is verified by applying to the evolved systems an additional local field $h'_i=\pm 2$ at each site $i$ and estimating the effect $\Delta_{h_i}\phi$ of this perturbation on the fitness $\phi(a)$: as seen in Fig.~\ref{fig:ex}, the sites responding to this local perturbation are the same as those participating to the conformational change. An overlap with sites sensitive to mutations ($\Delta_{a_i}\phi$) is also evident although less straightforward because a mutation at $i$ affects the couplings $J_{ij}(a_i,a_j)$ to neighboring sites in addition to the local field $h_i(a_i)$. In any case, a selective pressure for exquisite discrimination is sufficient in this model to generate with high probability distant allosteric sites and long-range mutational effects.

To assess the implications for adaptation, we define a measure of evolvability $\varepsilon(a)$ that reports the phenotypic diversity of sequences differing from $a$ by a single mutations. Here, the relevant phenotype is the set of ligands that a sequence can discriminate. In the limit where $\ell_r$ and $\ell_w$ are very similar, $|\ell_r-\ell_w|\to 0$, there is typically either no or a single value of $\ell_r$ for which $\phi(a)>0$. This value $\ell_r(a)$ provides a synthetic characterization of the phenotype of $a$. A measure $\varepsilon(a)$ of the phenotypic diversity of the neighborhood of $a$ can thus be defined from the number of different values that $\ell_r(a')$ takes when considering all the single mutants $a'$ of $a$. {\col In practice, we fix an interval of phenotypes of interest $[\ell_{\rm min},\ell_{\rm max}]=[-1,4]$, partition it into small subintervals of length $\delta=0.1$, and for each single mutant $a'$ of a given sequence $a$ find, if it exists, the interval to which $\ell_r(a')$ belongs: $\varepsilon(a)$ is then defined as the fraction of subintervals covered by the $(q-1)W(L+1)$ single-point mutants of $a$.} As seen in Fig.~\ref{fig:Nheff}B, $\varepsilon(a)$ correlates with $\bar\omega_r(a)$, consistent with the proposition that an extended conformational switch favors adaptation to new selective pressures; {\col this correlation does not depend on the choice of $\delta$ to resolve different phenotypes (Fig.~S\ref{FigS:delta})}. In contrast, $\varepsilon(a)$ does not correlate with $\phi(a)$ (Fig.~S\ref{FigS:w_phi}). 
{\cl Importantly, while $\varepsilon(a)$ is defined from the effect of single-point mutations only, it correlates with the capacity of a system to adapt to a change of selective pressure over multiple generations (Fig.~S\ref{FigS:change}).}

In summary, the two-dimensional spin model illustrates how selection for exquisite discrimination gives rise to a conformation switch, how this switch may involve a varying number of sites and how the potential for allostery and evolvability increases with the number of coupled sites that control the switch.

\section{Conservation, coevolution and thermal stability}

A relevant model of protein evolution should reproduce the salient statistical patterns present in multiple sequence alignments of natural protein families. Those families comprise sequences that evolved from a common ancestral sequence under presumably similar selective pressures. The simplest way to produce comparable alignments from the model is to take an evolved sequence as the ancestral sequence, generate from it $M$ independent trajectories, and collect the sequences at the end points into an alignment. Although this procedure assumes a trivial star-like phylogeny and strictly constant selection, we find it sufficient to produce statistical features comparable to those in natural alignments.

\begin{figure}[t]
\begin{center}
\includegraphics[width=\linewidth]{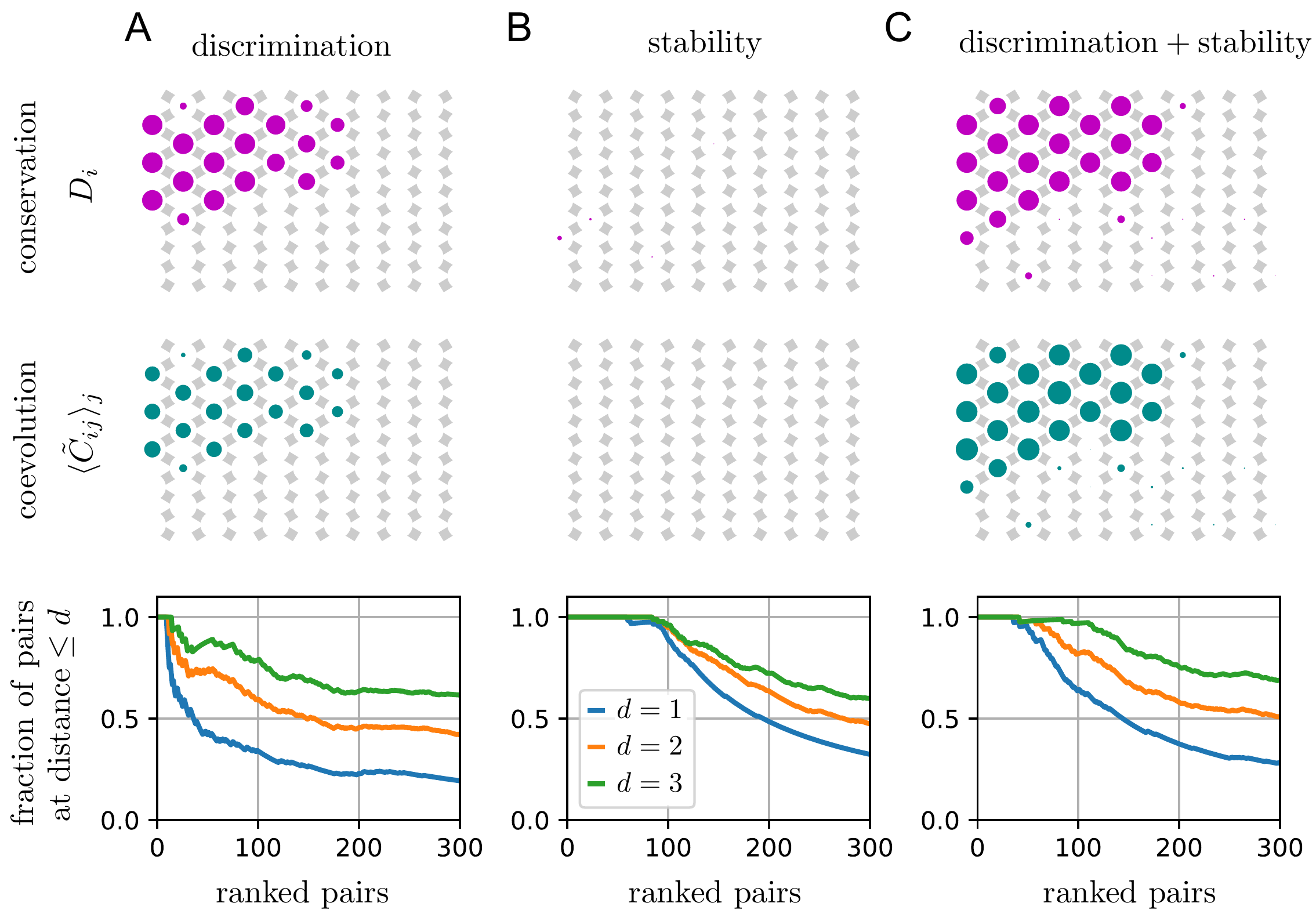}
\caption{Conservation and coevolution in multiple sequence alignments. Three alignments of 1000 sequences were generated from the same ancestral sequence, corresponding to the second column of Fig.~\ref{fig:ex}, but under different selective pressures: {\bf A.}~exquisite discrimination alone, {\bf B.}~stability alone, {\bf C.}~the two selections jointly. The first row indicates how the measure of evolutionary conservation $D_i$ defined in Eq.~\eqref{eq:Di} varies from site to site. {\col The second row indicates the mean value of the SCA correlation matrix $\langle \tilde C_{ij}\rangle_j$ for each site $i$, a quantity that combines conservation and coevolution~\cite{Rivoire:2016bl}.} The third row displays the performance of contact prediction by plmDCA~\cite{Ekeberg:2013fq}: pairs of sites are ranked from most to least coevolving and for the top $r$ pairs, the fraction in contact in the cylindrical structure is reported as a function of $r$, with contact defined either as distance $d=1$ (in blue), $d=2$ (orange) or $d=3$ (green). Selection for discrimination produces a set of evolutionary conserved {\col and correlated} sites that correspond to sites showing a strong mutational effect in the ancestral sequence (2nd column, 3rd row of Fig.~\ref{fig:ex}), but only few coevolving contacts. Selection for stability produces a different pattern, without strongly conserved sites but with many coevolving contacts. Finally, joint selection for discrimination and stability generates both a set of conserved sites and many coevolving contacts.\label{fig:alg}}
\end{center} 
\end{figure}

Given an alignment, a degree of evolutionary conservation can be defined at each site $i$ by the relative entropy
\beq\label{eq:Di}
D_i=\sum_{a_i=1}^qf_i(a_i)\ln\frac{f_i(a_i)}{q^{-1}},
\eeq
 where $f_i(a_i)$ is the frequency at which amino acid $a$ is present at position $i$ in the alignment. The pattern of evolutionary conservation $D_i$ essentially reproduces the patterns of conformational changes $\omega_{i,r}$ and allosteric effects $\Delta_{h_i}\phi$ in the ancestral sequence (Fig.~\ref{fig:alg}A, to be compared with the second column of Fig.~\ref{fig:ex}, which characterizes the ancestral sequence from which the alignment is generated). 

 Beyond conservation at individual positions, we can also analyze coevolution between sites through the correlation matrix $C_{ij}(a_i,b_j)=f_{ij}(a_i,b_j)-f_i(a_i)f_j(b_j)$ where $f_{ij}(a_i,b_j)$ is the joint frequency of $(a_i,b_j)$ at sites $i$ and $j$. Different statistical patterns can be extracted from $C_{ij}(a_i,b_j)$~\cite{Rivoire:2013cy,Cocco:2013el}. The statistical coupling analysis (SCA) thus extracts conserved global modes~\cite{Rivoire:2016bl} while the direct coupling analysis (DCA) extracts pairs of strongly coupled sites~\cite{Morcos:2011jg}. Applied to natural sequence alignments, SCA identifies groups of evolutionary conserved and coevolving sites called sectors, which are found to be structurally connected and associated with core functions of proteins~\cite{Rivoire:2016bl}. DCA, on the other hand, infers a ranked list of coevolving pairs, the top ones of which are found to be in contact in the three-dimensional structure~\cite{Morcos:2011jg}. Applying SCA to alignments generated from the model, we recover as a sector essentially the same set of sites that participate to the conformational switch and are identified from evolutionary conservation (Fig.~\ref{fig:alg}), comparable to what is obtained with natural alignments of proteins with a single sector. {\cl SCA combines conservation and correlations and here the sector is mostly controlled by conservation. In alignments of natural proteins, sequences of different specificities are often present which enhance the correlations.} Applying DCA yields top coevolving pairs that are in structural contact (Fig.~\ref{fig:alg}B); their number, however, is small compared to what is obtained from natural alignments of similar size and diversity.

One addition to the model is sufficient to correct for this discrepancy: a selective constraint on thermal stability. Our model indeed assumes that all sequences are folded into the same cylindrical shape without considering the possibility for some mutations to destabilize this structure. A simple way to integrate this possibility without explicitly modeling folding is to impose a maximal value $F^*$ for the free energy $F(a,\ell_0)$: above this value, the system is considered unfolded. To account for the fact that natural proteins are only marginally stable~\cite{Taverna:2002gj}, we choose $F^*$ well below the typical values of $F(a,\ell_0)$ in absence of stability constraint (Fig. S\ref{FigS:stab_phi}) but well above what may be obtained by minimizing $F(a,\ell_0)$ (Fig.~S\ref{FigS:stab_opt}). As shown in Fig.~\ref{fig:alg}B, the constraint $F(a,\ell_0)\leq F^*$ alone is sufficient to generate a large number of coevolving contacts, irrespectively of the exact value of $F^*$ (Fig.~S\ref{FigS:Fstar}). Imposing jointly the two selective pressures, exquisite discrimination and thermal stability, reproduces both features of natural alignments, a localized and evolutionarily conserved sector controlling binding affinity and specificity, and a large number of coevolving and contacting pairs of sites distributed across the structure (Fig.~\ref{fig:alg}C and Fig.~S\ref{FigS:pairs_D}).

\section{Discussion}

{\col Proteins derive their many functions from a few key properties, amongst which the capacity to stably fold into three-dimensional structures (stability), to selectively bind distinct ligands and substrates (specificity), to be regulated through long-range effects (allostery) and to adapt to changing selective pressures (evolvability). These different properties have been characterized in a number of instances but understanding how they are encoded into amino acid sequences remains a challenge. One puzzling feature is the ubiquity of long-range effects: mutational and evolutionary studies concur to indicate that binding and catalysis are affected by substitutions of amino acids more than 10\ \AA\ from the active site~\cite{licata1995long}. Another puzzling feature is the co-existence of two types of coevolution patterns in multiple sequence alignments: structurally connected groups of coevolving amino acids called sectors~\cite{Lockless:1999uf,Halabi:2009jc} as well as a variety of more isolated pairs of coevolving amino acids in contact in the three-dimensional structure~\cite{Morcos:2011jg}. Beyond case-by-case descriptions and statistical modeling, what explains the mutational and evolutionary patterns that we observe in protein sequences? To what extent are the different key properties of proteins independent of each other?  Without necessarily seeking to retrace natural history, what parsimonious evolutionary scenarios may generate comparable sequences and phenotypes?

Here, we proposed and analyzed such a scenario, which is based on one primary selective pressure, the requirement to achieve fine molecular recognition. First, we showed in a variety of different physical contexts, how discriminative binding requires a finely tuned conformational switch. Second, we argued that fine tuning may involve an extended subpart of the system to benefit from an enlarged evolutionary space. We then showed that extended sectors indeed emerge in a physical model with short range interactions where selection operates only locally. Third, we showed that the sensitivity of this sector to perturbations implies a potential to evolve allosteric regulation and to adapt to new selective pressures. The model explains both the conformation changes observed in protein structures and the coevolving sectors observed in protein sequences. In this scenario, adaptability is not opposed to high-performance but comes as its natural consequence. The arguments are partly independent of the physical nature of the system, which only needs to be a structured network with short-range interactions subject to evolution. We therefore chose to illustrate the scenario with one of the simplest physical models, a spin model. Finally, we showed how the second type of coevolution patterns, distributed pairs of coevolving contacts, can be explained within the same framework by taking into account an additional selective pressure, thermal stability. Our evolutionary scenario is parsimonious in the sense that only two selective pressures, discriminative binding and thermal stability, are invoked, with long-range effects arising as a by-product.}

{\col This is in sharp contrast with several physical models recently been proposed for the evolution of allostery, which are based on a direct selection for long-range regulation~\cite{Hemery:2015ei,flechsig2017design,rocks2017designing,tlusty2017physical,yan2017architecture}. A general evolutionary argument and specific experiments, however, argue strongly against a direct selection for long-range effects in the initial steps of the evolution of allosteric regulation.}
\textcolor{black}{From an evolutionary perspective, the emergence of allostery by direct selection for regulation is {\col indeed considered} implausible {\col in the biological literature } as it assumes the concerted evolution of two proteins, one becoming regulated and the other one regulating it~\cite{Kuriyan:2007}. As for the evolution of other complex systems, the conundrum disappears if considering the pre-existence of components and properties that evolved under unrelated selective pressures. The evolution of allosteric regulation is{\col , instead,} easily explained if long-range effects pre-exist any direct selective pressure to evolve them. The proposal that proteins have an inherent propensity for allostery has been made previously~\cite{Gunasekaran:2004} and {\col is strongly supported by } a number of experimental results, starting with the repeated discovery of serendipitous allosteric sites with non-physiological effectors~\cite{Hardy:2004gy,zorn:2010}. Latent allostery is also demonstrated by the successful engineering of allosteric regulation at several surface sites of a protein~\cite{Lee:2008gd,Reynolds:2011gs,pincus2017evolution}; these sites were not known as regulatory targets but, consistent with our model, they are evolutionarily linked through a sector to the active site. These works, and more generally the ubiquity of long-range effects in proteins~\cite{licata1995long,Lockless:1999uf,Daugherty:2000gz,Morley:2005kc}, unequivocaly reveal the prevalence of latent allostery. {\col Finally, compelling evidence for pre-existing long-range effects as precursors of allosteric regulation is provided by an experimental study of MAP kinases~\cite{Coyle:2013dn}: }an effector $A$ allosterically regulating a kinase $K$ in one species of yeast was shown to regulate evolutionary related kinases $K'$ in other species where no such regulation or analog to $A$ had evolved, {\col in line with a scenario where} that $A$ had an allosteric effect on the common ancestor of $K$ and $K'$ prior to any selection for allosteric regulation. While convincingly indicating latent allostery as an origin of allosteric regulation, these past studies do not explain, however, how latent allostery arises in the first place. Our model is the first, to our knowledge, to provide a possible explanation.}

{\col Our evolutionary scenario also contrasts with other explanations for the origin of evolvability}. 
In some {\col previous} models, flexibility is linked to evolvability, but evolution under constant selective pressure leads to a phenomenon of canalization where flexibility and evolvability both become increasingly limited~\cite{Ancel:2000gj}. This is resolved in other models by considering an evolutionary history of {\col temporally varying } selecting pressures {\col where, for instance, the nature of the ligand changes in the course of evolution.}~\cite{Parter:2008bb,Hemery:2015ei,Raman:2016gv}. While such fluctuations are likely to be relevant to protein evolution, our model shows that they are not necessary to generate evolvable proteins. The two factors, selection for exquisite discrimination and fluctuating selective pressures, are however non-exclusive and may reinforce each other. In our model, a fluctuating selection may thus contribute to select for systems with more extended conformational switches. The two factors may in fact reflect the same principle: the presence of related but partly conflicting constraints. The relationship between these constraints may be more critical than their simultaneous or successive occurrence. 

{\cl The contribution of conformational changes to molecular recognition has also been explained by a different mechanism called conformational proofreading~\cite{savir2007conformational}. In this model, a structural mismatch between the protein and its substrates favors correct recognition by penalizing binding to the wrong ligand more than it does to the right ligand. Here, we considered a stronger notion of molecular recognition called exquisite discrimination where binding to the wrong ligand is less favorable than binding to the solvent in addition to being less favorable than binding to the right ligand. We showed exquisite discrimination to be achievable through a particular type of conformational change, which can in particular take the form of a switch between two local minima of the potential (Fig.~1A). While conformational proofreading is particularly relevant to proteins involved in search processes~\cite{vlaminck}, exquisite discrimination may be more relevant to enzymes, which have to bind to a reactant but release its product. Despite differences, the two mechanisms imply a similar notion of fine-tuning, which we propose here to be a sufficient factor for evolving an extended set of coupled sites.}

{\col By examining the implications of discriminative recognition, our model complements the many studies focused on the folding and thermal stability of proteins. This includes models that analyzed} the interplay between binding affinity and thermal stability~\cite{Miller:1997es,Williams:2001ua,Manhart:2015eg}. Our model illustrates a simple dichotomy. On one hand, selection for exquisite discrimination leads to a conserved set of coevolving positions structurally connected to the binding site, analogous to the sectors inferred from multiple sequence alignment of natural proteins~\cite{Rivoire:2016bl}. On the other hand, constraints on thermal stability lead to a large number of contacting coevolving sites distributed across the structure, analogous again to what is found in natural proteins~\cite{Morcos:2011jg}.  {\cl Consistent with this scenario, the method that best infers structural contacts from coevolution (DCA) is also very effective at scoring sequences for their thermal stability~\cite{Morcos14,Figliuzzi15}.} Our model treats unfolded conformations implicitly but the generation of coevolving contacts from selection for thermal stability was previously demonstrated in lattice protein models where unfolded conformations are explicitly considered~\cite{Jacquin:2016cl}. At the origin of the two distinct statistical signatures are two essential differences between binding and stability constraints. Physically, binding involves a localized interaction while stability is a structurally distributed property. Evolutionarily, selective pressures on stability are typically less stringent than selective pressures on binding, as reflected by the marginal stability of most globular proteins~\cite{Taverna:2002gj} and the scarcity of mutations improving binding in wild-type proteins~\cite{McLaughlinJr:2012hw}. As a result, the stability problem has high entropy in sequence space, with different solutions involving different isolated interactions, while the binding problem has low entropy, with solutions all involving essentially the same clustered interactions.

{\col Partial or global unfolding induced by perturbations at sites contributing to thermal stability, which may be located arbitrarily far from the active site, is in fact another mechanism by which protein activity may respond to distant perturbations without any explicit selection for allosteric regulation.} Allosteric switches based on partial or global unfolding are indeed documented in several proteins~\cite{Motlagh:2014kc} and can provide a mechanism for adaptation~\cite{Saavedra:2018ea}. For such mechanism, no evolutionarily conserved pathway linking the allosteric site to the active site is necessary. 

{\col Thermal fluctuations around a folded state also provide a way by which perturbations may propagate through long distances, possibly without major conformational change~\cite{Cooper:1984cn}. This mechanism is evidenced in some proteins~\cite{Popovych:2006} and is explained by elastic inhomogeneities~\cite{mcleish:2013}. As these inhomogeneities are generic features of proteins, their presence may also precede any selection for allosteric regulation. Following our approach, it would be interesting to study this other scenario, taking into account the demonstrated relation to coevolving contacts~\cite{Townsend:2015}. This scenario may depend on the nature of the physical interactions since, at least in some proteins, fluctuation-based allostery operates through elastic motions~\cite{Townsend:2015}. Spin models, where only a few discrete states are accessible at each site, may instead be more suited to a description of side-chain fluctuations between a few rotameric states attached to a fixed backbone, which dominate fluctuations in some other proteins~\cite{dubay2011long}. Going beyond the generic effects analyzed in this work to study evolutionary scenarios that rely on more specific physical mechanisms is a natural avenue for future work.}

{\col While other evolutionary scenarios are possible and should be considered, the present scenario can already be confronted to experimental tests. First, it is possible to analyze} the physical underpinning of the different coevolution patterns. {\col Our model predicts that mutating sector positions impacts function (binding, catalysis) and possibly also thermal stability, while mutating coevolving contacts outside of sectors impacts mostly thermal stability.} The idea that multiple sequence alignments contain correlations of different nature is already a key principle behind the statistical coupling analysis (SCA), where the most conserved correlations are up-weighted to highlight functional coevolution~\cite{Lockless:1999uf,Rivoire:2016bl}. This approach is supported by several experimental studies, including the design of functional proteins by reproducing the patterns of functional coevolution~\cite{Socolich:2005js,Russ:2005kc}. The same principles may be extended to design sequences enhancing or ignoring other types of correlations found in multiple sequence alignments. For instance, in the context of the direct coupling analysis (DCA), which also provides a generative model~\cite{figliuzzi2018pairwise}, we may predict that designing sequences at low statistical temperature will lead to more stable proteins.

{\col A direct experimental test of the central idea that a selection for exquisite discrimination is sufficient to generate an extended sector is also possible. This can be done} in two steps. First, proteins can be evolved to have desired specificities, using methods of directed evolution {\color{black} such as phage or yeast display~\cite{Smith:1997vn,Levin:2006}. These methods allow for the selection of proteins with high affinity to an arbitrary target ligand from populations of billions of different variants of a protein. This selection can be repeated with intervening steps of mutations and amplification, thus emulating {\it in vitro} evolution by natural selection. Specificity can be controlled by combining positive selection for binding to immobilized targets with negative selection for not binding to soluble targets that are washed away. The approach is powerful enough to generate proteins specific to a particular conformational state of a target protein~\cite{Nizak:2003} or, on the contrary, cross-reactive to a range of variants of a target protein~\cite{Garcia:2010}.}  Second, the presence of long-range effects can be assayed by screening and sequencing mutants of the evolved proteins~\cite{Fowler:2014gq}.  {\color{black} Starting from an artificially evolved protein, a population of mutants can be produced and selected for specific binding: using high-throughput sequencing, the frequency of each mutant before and after selection can be known, which reveals which mutations have an effect on binding. Our prediction is that proteins evolved to be highly specific will exhibit more deleterious mutations far from their binding site. As specificity can be finely controlled during selection and as sensitivity to mutations can be assayed quantitatively and mapped to three-dimensional structures within a few angstroms, these experiments can not only test the prediction of our toy model but provide data for elaborating more accurate models.}  

\hspace{15cm}

\ \\

{\small 
\begin{acknowledgments}
This work benefited from multiple inspiring discussions with Rama Ranganathan, Cl\'ement Nizak and Anton Zadorin, and from comments by Yaakov Kleeorin. It was supported by FRM AJE20160635870.
\end{acknowledgments}
}

\clearpage
\newpage

\appendix

\makeatletter
\makeatletter \renewcommand{\fnum@figure}
{\figurename~S\thefigure}
\makeatother
\setcounter{figure}{0}

\renewcommand\theequation{S\arabic{equation}}
\setcounter{equation}{0}

\begin{center}
{\bf APPENDIX}
\end{center}

\section*{A. Fitness function for exquisite discrimination}\label{sec:tradeoff}

Two ligands $\ell_r$ and $\ell_w$ with binding free energies $\Delta F_r(a)$ and $\Delta F_w(a)$ and chemical potentials $\mu_r$ and $\mu_w$ have, at thermal equilibrium, probabilities $p_r(a)$ and $p_w(a)$ to bind, with
\beq
p_k(a)=\frac{e^{-\beta(\Delta F_k(a)-\mu_k)}}{1+e^{-\beta(\Delta F_r(a)-\mu_r)}+e^{-\beta(\Delta F_w(a)-\mu_w)}},
\eeq
where $k=r$ or $w$.
If binding to $\ell_r$ is desirable and binding to $\ell_w$ is not, the sequence $a$ has to maximize $p_r(a)$ and minimize $p_w(a)$. The problem may be formalized as the optimization of a fitness function $\phi(a)$ {\col that is a decreasing function of $\Delta F_r(a)$ and an increasing function of $\Delta F_w(a)$.

The choice of $\phi(a)=\min(-\Delta F_r,(a),\Delta F_w(a))$ leads to one of the simplest fitness functions that captures this trade-off. Imposing $\phi(a)>0$ is equivalent to imposing $F(a,\ell_r)<F(a,\ell_0)<F(a,\ell_w)$. As another important factor controlling the difficulty of discrimination is the relative concentration of the two ligands. A natural generalization is therefore $\phi(a)=\min(-\Delta F_r(a)+\mu,\Delta F_w(a)-\mu)$, so that $\phi(a)>0$ is equivalent to imposing $F(a,\ell_r)-\mu<F(a,\ell_0)<F(a,\ell_w)-\mu$. We show in Fig.~S\ref{FigS:fit} that taking $\mu=-1$ or $\mu=+1$ gives similar results than $\mu=0$. 

To further demonstrate the robustness of the results to the choice of the fitness function, we also show that the less physical choice of $\phi(a)=-\Delta F_r(a)\times\Delta F_w(a)$, corresponding to the requirement that $\Delta F_r,(a)$ and $\Delta F_w(a)$ should both have large absolute values but opposite signs, also gives similar results (Fig.~S\ref{FigS:fit}).}

\section*{B. Unidimensional models}

The potentials $U(x,a,\ell)=U(x,h=\ell-a)$ of the models presented in Fig.~\ref{fig:elementary} have a common symmetry: 
\beq
U(x,-h)=U(-x,h). 
\eeq
The free energy $F(h)=-\beta^{-1}\ln\int dx\ e^{-\beta U(x,h)}$
 is therefore symmetric around $h=0$: $F(h)=F(-h)$, where $F(h)$ is a decreasing function of $h$ for $h>0$. Given $\ell_0,\ell_r,\ell_w$, three different cases thus arise when considering the existence of $a$ satisfying the condition $F(a,\ell_r)<F(a,\ell_0)<F(a,\ell_w)$:

(i) if $\ell_w<\ell_0<\ell_r$ (respectively, $\ell_r<\ell_0<\ell_w$), the condition is satisfied by taking any $a<\ell_w$ (resp. any $a>\ell_w$) so that $\ell_w-a,\ell_0-a,\ell_r-a$ are of same sign.

(ii) if $\ell_0<\ell_w<\ell_r$ (respectively, $\ell_r<\ell_w<\ell_0$), the condition is satisfied by taking $\ell_w-a<-\ell_0+a<\ell_r-a$, i.e. $(\ell_0+\ell_w)/2<a<(\ell_0+\ell_r)/2$ (resp. $\ell_w-a<-\ell_0+a<\ell_r-a$, i.e. $(\ell_0+\ell_r)/2<a<(\ell_0+\ell_w)/2$) so that $\ell_w-a,\ell_r-a$ on one hand and $\ell_0-a$ on the other are of opposite signs.

(iii) if $\ell_0<\ell_r<\ell_w$ (respectively, $\ell_w<\ell_r<\ell_0$), the condition cannot be satisfied.

As $\langle x\rangle_{-h}=-\langle x\rangle_h$, (ii) involves a change of sign for the mean conformation while (i) does not. For the models of Fig.~\ref{fig:elementary}, solutions to (i) in fact exist that have arbitrarily small relative conformational changes, $(\langle x\rangle_{h_r}-\langle x\rangle_{h_0})/\langle x\rangle_{h_0}\to 0$, obtained for $a\to\pm\infty$.

The problem of exquisite discrimination is to be compared to the problem of maximizing the affinity to a single target $\ell_r$. As $\partial^2F(h)/\partial h^2=-(\langle x^2\rangle_h-\langle x\rangle_h^2)\leq 0$, $F(h)$ is concave. In general, it is even strictly concave, the single-spin model at zero temperature where $F(h)=-|h|$ being a degenerate case. This implies that $F(a,\ell_r)-F(a,\ell_0)$ is minimized by taking $a\to+\infty$ when $\ell_r>\ell_0$ and $a\to-\infty$ when $\ell_r<\ell_0$. More generally, the condition $F(a,\ell_r)<F(a,\ell_0)$ may be seen as a limit of the condition $F(a,\ell_r)<F(a,\ell_0)<F(a,\ell_w)$ when ${\rm sign}(\ell_0)(\ell_0-\ell_w)\to 0^+$, corresponding to case (i).

\subsubsection*{Single-spring elastic network}\label{sec:sol1}

The model of Fig.~\ref{fig:elementary}A is defined by 
\beq\label{eq:U1}
U(x,h)=\frac{1}{2}k(|x|-r)^2-hx,\qquad x\in\mathbb{R},
\eeq
which verifies $U(x,-h)=U(-x,h)$. Here, $k>0$ represents the stiffness of the spring and $r>0$ its equilibrium length. While it is possible to write a general analytical formula for the free energy $F(h)$ at any $\beta$, it is more illuminating to consider the zero-temperature limit $\beta\to\infty$. In this limit, $F(h)=\min_x U(x,h)$ is reached for $x=\hat x(h)$ with
\beq
F(h)=-\frac{1}{2}\frac{h^2}{k}-|h|r,\qquad \hat x(h)=\frac{h}{k}+{\rm sign}(h)r.
\eeq
$F(h)$ and $\hat x(h)$ are represented in Fig.~\ref{fig:elementary}A for $r=1$, $k=2$, $h_0=-1/2$, $h_w=1/4$, $h_r=3/4$.

\subsubsection*{Two-spring elastic network}\label{sec:sol2}

The model of Fig.~\ref{fig:elementary}B is defined by 
\beq\label{eq:U2}
U(x,h)=k\left(\sqrt{x^2+d^2}-r\right)^2-hx,\qquad x\in\mathbb{R},
\eeq 
which verifies $U(x,-h)=U(-x,h)$. Here, $k>0$ represents the common stiffness of the two springs, $r>0$ their common equilibrium length and $2d>0$ the distance between the two fixed points at which they are attached. In the zero temperature limit $\beta\to\infty$, the free energy is the minimum of $U(x,h)$, reached for $\hat x(h)$ solution to
\beq
\hat x(h)\left(1-\frac{r}{\sqrt{\hat x(h)^2+d^2}}\right)=\frac{h}{2k}.
\eeq
In Fig.~\ref{fig:elementary}B, the graph of $F(h)$ versus $h$ is obtained by parametrizing $F(h)$ and $h$ by $\hat x$: $F(h)=U(\hat x,h(\hat x))$ and $h(\hat x)=2k \hat x(1-r(\hat x^2+d^2)^{-1/2})$. Fig.~\ref{fig:elementary}B corresponds to $k=1$, $r=1$, $h_0=-1/2$, $h_w=1/4$, $h_r=3/4$. Depending on the value of $r$, $U(x,h)$ may have either one or two local minima (Fig.~S\ref{FigS:spring_var}).

\subsubsection*{Single-node harmonic model}\label{sec:solG}

A single-node Gaussian elastic network model is defined by 
\beq\label{eq:UG}
U(x,h)=\frac{1}{2}k(x-r)^2-hx,\qquad x\in\mathbb{R}.
\eeq
It verifies $U(x,-h)=U(-x,h)$ only for $r=0$ but the model is physically equivalent to a model with potential $V(x,h)=U(x+r,h)+hr$ which verifies $V(x,-h)=V(-x,h)$ for any $r$. The conclusions are therefore the same as for the previous models. The free energy is 
\beq
F(h)=-\frac{h^2}{2k}-hr+\frac{1}{2\beta}\ln\frac{\beta k}{2\pi}.
\eeq
It corresponds graphically to an inverted parabola, qualitatively similar to $F(h)$ for the models of Fig.~\ref{fig:elementary}.

\subsubsection*{Shape space model}

In shape space models~\cite{Perelson:1979gb}, no flexibility is considered and the conformations $x$ are formally identified with the sequences $a$. The sequences $a$ and ligands $\ell$ are points in a $d$-dimensional Euclidean space and the binding energy between $a$ and $\ell$ is a function of their Euclidean distance, in the simplest case $F(a,\ell)=\|a-\ell\|_2$. In dimension $d=1$, this corresponds to $F(h)=|h|$ and exquisite discrimination is possible when $\ell_0<\ell_r<\ell_w$ provided $(\ell_0+\ell_r)/2<a<(\ell_0+\ell_w)/2$, similar to the criterion derived for elastic models.

\subsubsection*{Single-spin model}\label{sec:sols}

The model of Fig.~\ref{fig:elementary}C is defined by 
\beq
U(x,h)=-hx,\qquad x\in\{-1,+1\},
\eeq
which verifies $U(x,-h)=U(-x,h)$.
The free energy is
\beq\label{eq:Fhspin}
F(h)=-\beta^{-1}\ln\left(e^{\beta h}+e^{-\beta h}\right).
\eeq
In the zero-temperature limit $\beta\to\infty$, $F(h)=-|h|$, which is identical up to the sign to the free energy of a one-dimensional shape space model.

Let assume $\ell_0=0$ and $\ell_r>0$ to analyze the nature of the solutions satisfying $F_r<F_0<F_w$ where $F_r=F(a,\ell_r)$, $F_0=F(a,\ell_0)$ and $F_w=F(a,\ell_w)$. We proceed by examining the different possible cases:

1. if $a>0$, then $F_0=-a$ and $F_r=-a-\ell_r$ satisfies $F_r<F_0$.

\tab  1.1. if $a<-\ell_w$, then $F_w=a+\ell_w$ and $F_0<F_w$ implies $-a<a+\ell_w$, i.e., $-\ell_w/2<a$. In total, $-\ell_w/2<a<-\ell_w$. This is possible provided $\ell_w<0$. In this case, $F_r-F_0=-\ell_r<0$ and $F_w-F_0=\ell_w+2a\leq -\ell_w$, whose maximum is reached for $\hat a=-\ell_w$, yielding $F_w-F_0=-\ell_w>0$. A solution exists in this case.

\tab 1.2. if $-\ell_w<a$, then $F_w=-a-\ell_w$ and $F_0<F_w$ implies $-a<-a-\ell_w$, i.e., $\ell_w<0$. 
In total, $0<-\ell_w<a$. This is possible provided $\ell_w<0$. In this case, $F_r-F_0=-\ell_r<0$, $F_w-F_0=-\ell_w>0$ independent of the value of $a$. A solution exists in this case.

2. if $a<0$, then $F_0=a$.

\tab 2.1. if $a<-\ell_r$, $F_r=a+\ell_r$ and $F_r<F_0$ implies $a+\ell_r<a$, which is inconsistent with the assumption that $\ell_r>0$. No solution exists in this case.

\tab 2.2. if $-\ell_r<a$, $F_r=-a-\ell_r$ and $F_r<F_0$ implies $-a-\ell_r<a$, i.e., $-\ell_r/2<a$.

\tab\tab 2.2.1. if $a<-\ell_w$, $F_w=a+\ell_w$ and $F_0<F_w$ implies $a<a+\ell_w$, i.e., $\ell_w>0$. In total, $-\ell_r/2<a<-\ell_w<0$. This is possible provided $0<\ell_w<\ell_r/2$. In this case, $F_0-F_r=2a+\ell_r\leq \ell_r-2\ell_w$, $F_w-F_0=\ell_w$ with maximum reached for $\hat a=-\ell_w$.

\tab\tab 2.2.2. if $-\ell_w<a$, $F_w=-a-\ell_w$ and $F_0<F_w$ implies $a<-a-\ell_w$, i.e., $a<-\ell_w/2$. In total, $\max(-\ell_r/2,-\ell_w)<a<-\ell_w/2$. Possible provided $0<\ell_w<\ell_r$. $F_0-F_r=2a+\ell_r\leq \ell_r-\ell_w$, $F_w-F_0=-2a-\ell_w\leq \min(\ell_r-\ell_w,\ell_w)$, which cannot be both maximized. If maximizing $\min(F_0-F_r,F_w-F_0)$, the optimum is for $\hat a=-(\ell_r+\ell_w)/2$ in which case $F_0-F_r=F_w-F _0=(\ell_r-\ell_w)/2$.
 
All together, the nature of the solution depends on the ratio $\ell_w/\ell_r$:

\tab -- if $\ell_w/\ell_r<0$: two solutions, $\hat x_0=+$, $\hat x_r=+$, $\hat x_w=\pm$, with $\hat x_w=+$ for the solution optimizing $\phi(a)$;

\tab -- if $0<\ell_w/\ell_r<1/2$: two solutions $\hat x_0=-$, $\hat x_r=+$, $\hat x_w=\pm$, with $\hat x_w=-$ for the solution optimizing $\phi(a)$;

\tab -- if $1/2<\ell_w/\ell_r<1$: one solution $\hat x_0=-$, $\hat x_r=+$, $\hat x_w=+$; 

\tab -- $1<\ell_w/\ell_r$: no solution.

A switch is thus involved whenever $\ell_0<\ell_w<\ell_r$.

\section*{C. Two-dimensional spin model}

Given fields $h_i(a_i)$ and couplings $J_{ij}(a_i,a_j)$ defined respectively on each node $i$ and each edge $ij$ of the lattice represented in Fig.~\ref{fig:scheme}A, and given an external field $\ell$ applied to the red node $b$ in Fig.~\ref{fig:scheme}A, the free energy $F(a,\ell)$ is computed exactly from Eq.~\eqref{eq:Uxal} by transfer matrices using free boundary conditions for the spins at the left and right extremities. Performing the calculation from right to left allows for an efficient evaluation of $F(a,\ell)$ for different values of $\ell$ since in this case changing $\ell$ affects only the last layer.

The capacity of a system to discriminate between two ligands $\ell_r$ and $\ell_w$ is scored by $\phi(a)$ given by Eq.~\eqref{eq:phia}. To evolve systems under a selection for discrimination, a Metropolis Monte Carlo procedure is followed. It starts from a random sequence and generates a trajectory by iterating $T$ times the following steps: a site $i$ is chosen at random, an $a_i'\neq a_i$ is chosen at random between the $q-1$ possibilities, and the substitution is accepted if $r<\exp(\gamma(\phi(a')-\phi(a)))$ where $r$ is a random number uniformly drawn in $[0,1]$ and $a'$ is sequence $a$ with $a_i'$ substituted for $a_i$. The parameters $\gamma=100$ and $T=1000$ are taken. As shown in Fig.~S\ref{FigS:time}A, this number of iterations is sufficient for $\phi(a)$ to reach an equilibrium value.

This value is in most case close to the theoretical maximum that $\phi(a)$ may reach (Fig.~\ref{fig:Nheff}). This maximum is the same for the two-dimensional spin model and for the single-spin model. It is reached when $F(h_0=a+\ell_0)=(F(h_r=a+\ell_r)+F(h_w=a+\ell_w))/2$ with $F(h)$ given by Eq.~\eqref{eq:Fhspin}. With the parameters taken in the main text, $\max_a\phi(a)\simeq 0.41$.

The different systems in Fig.~\ref{fig:ex} correspond to different choices of the function $h_i(a_i)$ and $J_{ij}(a_i,a_j)$. Different solutions may also be obtained with the same mapping between sequences $a$ and fields $h_i(a_i)$ and couplings $J_{ij}(a_i,a_j)$ but different initial conditions or/and different series of proposed mutations. The system in the second column of Fig.~\ref{fig:ex} (with $\bar\omega_r=0.11$) is evolved under an additional constraint for thermal stability to serve as a common ancestral sequence for the alignments of Fig.~\ref{fig:alg}. 

\subsubsection*{Thermal stability}\label{sec:stab}

A system is considered thermally stable if its free energy is below a fixed threshold, $F(a,\ell_0)<F^*$. In Fig.~\ref{fig:alg}, the value $F^*=-350$ is chosen for being smaller than the free energy of systems evolved without stability constraints (Fig.~S\ref{FigS:stab_phi}, top left panel) but larger than the free energy of systems optimized for thermal stability (Fig.~S\ref{FigS:stab_opt}). 

To evolve stable and functional sequences, we first generate a stable sequence by a Metropolis Monte Carlo procedure where the scoring function is $-F_0(a)$ rather than $\phi(a)$ and stop the iterations as soon as the condition $F(a,\ell_0)<F^*$ is fulfilled. Starting from this stable sequence, we then generate a trajectory as before with the only modification that mutations causing $F(a,\ell_0)>F^*$ are systematically rejected (Fig.~S\ref{FigS:time_stab}). The systems thus produced are marginally stable with free energies $F(a,\ell_0)$ close to the threshold $F^*$ (Fig.~S\ref{FigS:stab_phi}).

\subsubsection*{Characterization}\label{sec:char}

The mean conformations $\langle x_i\rangle_{a,\ell_r}$ entering the definition of $\omega_{i,r}(a)$ in Eq.~\eqref{eq:omega} are computed exactly by transfer matrices. In Fig.~\ref{fig:Nheff}, only the fittest evolved systems with $\phi(a)>0.35$ are considered, which represent 94\% of 1000 evolved sequences. How $\bar\omega_r(a)$ relates to $\phi(a)$ is shown in Fig.~S\ref{FigS:w_phi}B.

To quantify evolvability in Fig.~\ref{fig:Nheff}B, we fix an interval of phenotypes of interest $[\ell_{\rm min},\ell_{\rm max}]=[-1,4]$, partition it into small subintervals of length $\delta=0.1$, and for each single mutant $a'$ of a given sequence $a$ find, if it exists, the interval to which $\ell_r(a')$ belongs. The fraction of subintervals covered by the $(q-1)W(L+1)$ single-point mutants of $a$ defines $\varepsilon(a)$, a measure of the phenotypic diversity its neighborhood in sequence space. How $\varepsilon(a)$ relates to $\phi(a)$ is shown in Fig.~S\ref{FigS:w_phi}D. {\col How the results depend on the choice of $\delta$ is shown in Fig.~S\ref{FigS:delta}.}

In Fig.~\ref{fig:ex}, the sizes of the red dots in A are proportional to $\omega_{i,r}$ defined in Eq.~\eqref{eq:omega}, the sizes of the green dots in B are proportional to $\Delta_{h_i}\phi$, defined for each site $i$ as the maximal decrease in fitness $\phi$ when adding either $h_i'=-2$ and $h_i'=+2$ to $h_i(a_i)$ at $i$, and the sizes of the blue dots in C are proportional to $\Delta_{a_i}\phi$, defined for each site $i$ as the mean loss in fitness when considering all possible $q-1$ mutations of $a_i$. The distribution of the effects of all single and double mutations are shown in Fig.~S\ref{FigS:time}B-C.

\subsubsection*{Multiple sequence alignments}

The three alignments of Fig.~\ref{fig:alg} are all generated from the same ancestral sequence, a stable and functional sequence that corresponds to the second column of Fig.~\ref{fig:ex} (Fig. S\ref{FigS:time_stab}). The $M=1000$ sequences in each alignment are obtained from independent trajectories starting from this sequence.

For the first alignment (labelled ``discrimination''), no constraints of stability is enforced (formally $F^*=\infty$) and selection is based on $\phi(a)$ given by Eq.~\eqref{eq:phia}. For the second alignment (``stability''), no constraint on discrimination is enforced (formally $\gamma=0$) and selection is limited to $F(a,\ell_0)<F^*$. For the third alignment (``discrimination + stability''), the two selective constraints are jointly imposed. The distributions of $F(a,\ell_0)$ and $\phi(a)$ in these three alignments are shown in Fig.~S\ref{FigS:stab_phi}. Taking a larger or smaller value of $F^*$ leads to similar conclusions (Fig.~S\ref{FigS:Fstar}).

\subsubsection*{Inference of coevolution}

Given an alignment, coevolving contacts are inferred by the pseudo-likelihood maximization method for direct coupling analysis~\cite{Ekeberg:2013fq_re} using an implementation in Julia~\cite{juliaDCA} with default parameters except for a reduced alphabet of $q=5$ amino acids. The algorithm outputs a ranked list of coevolving pairs. Contacts are defined based on the distance $d$ between nodes in the lattice of Fig.~\ref{fig:scheme}A: $d=1$ refers to two nodes connected by an edge and $d\geq 2$ to two nodes connected by $d$ edges through $d-1$ intermediate nodes.

For Fig.~\ref{fig:ex}, the SCA matrix $\tilde C_{ij}$ is calculated as described in Box 1 of Ref.~\cite{Rivoire:2016blre} without any preprocessing except for a regularization parameter $\lambda=0.1$. Background frequencies are taken to be $1/q$ for all $q=5$ amino acids. Fig.~S\ref{fig:ex} shows the mean values $\langle \tilde C_{ij}\rangle_j$ of each row $i$ of the matrix $\tilde C_{ij}$.

\clearpage

\newpage

\begin{widetext}

\begin{figure}[t]
\begin{center}
\includegraphics[width=.8\linewidth]{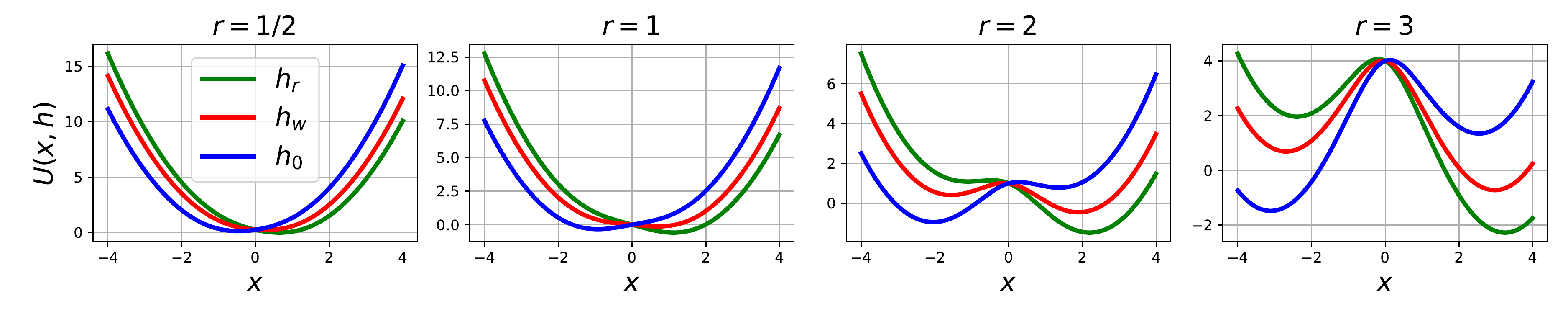}
\caption{Potential $U(x,h)$ for the model of Fig.~\ref{fig:elementary}B with springs of different equilibrium length $r$. $U(x,h)$ is given by \eqref{eq:U2} with here $k=1$, $d=1$, $h_0=-1/2$, $h_w=1/4$, $h_r=3/4$ such that $\min_xU(x,h_r)<\min_xU(x,h_0)<\min_xU(x,h_w)$, i.e., the global minimum of the blue curve is in-between the global minima of the red and green curves. For $r\leq 1$, $U(x,h)$ has a single minimum while for $r>1$ it has two minima. Fig.~\ref{fig:elementary}B corresponds to $r=1$. \label{FigS:spring_var}}
\end{center} 
\end{figure}

\begin{figure}[t]
\begin{center}
\includegraphics[width=.7\linewidth]{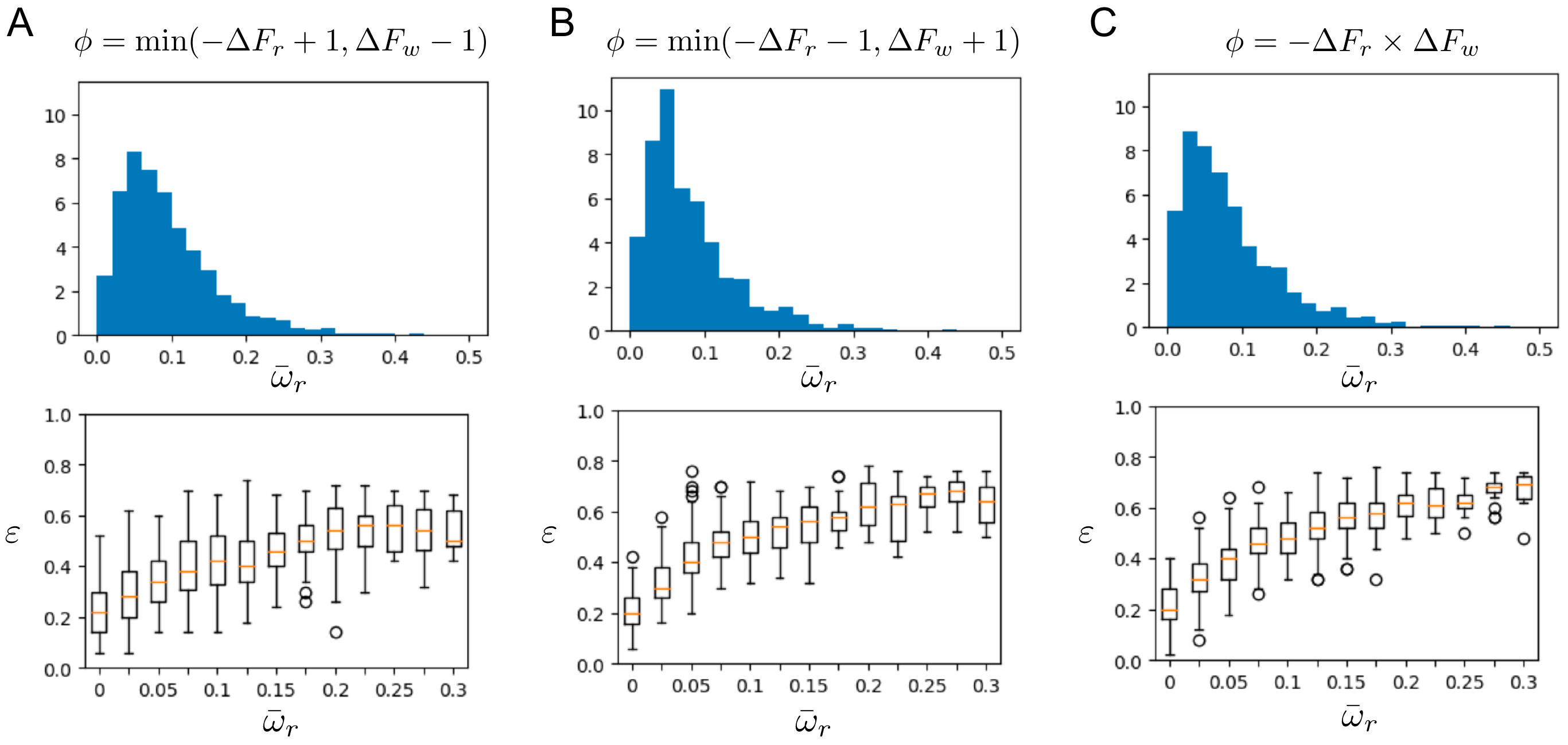}
\caption{{\col Graphs of Fig.~\ref{fig:Nheff} for alternative choices of the fitness function $\phi(a)$. {\bf A.} $\phi(a)=\min(-\Delta F_r(a)+\mu,\Delta F_w(a)-\mu)$ with $\mu=1$.  {\bf B.} $\phi(a)=\min(-\Delta F_r(a)+\mu,\Delta F_w(a)-\mu)$ with $\mu=-1$. {\bf C.} $\phi(a)=-\Delta F_r,(a)\times\Delta F_w(a)$. In any case, the results are qualitatively similar to those of Fig.~\ref{fig:Nheff}, which corresponds to $\phi(a)=\min(-\Delta F_r(a)+\mu,\Delta F_w(a)-\mu)$ with $\mu=0$.} \label{FigS:fit}}
\end{center} 
\end{figure}

\begin{figure}[t]
\begin{center}
\includegraphics[width=.5\linewidth]{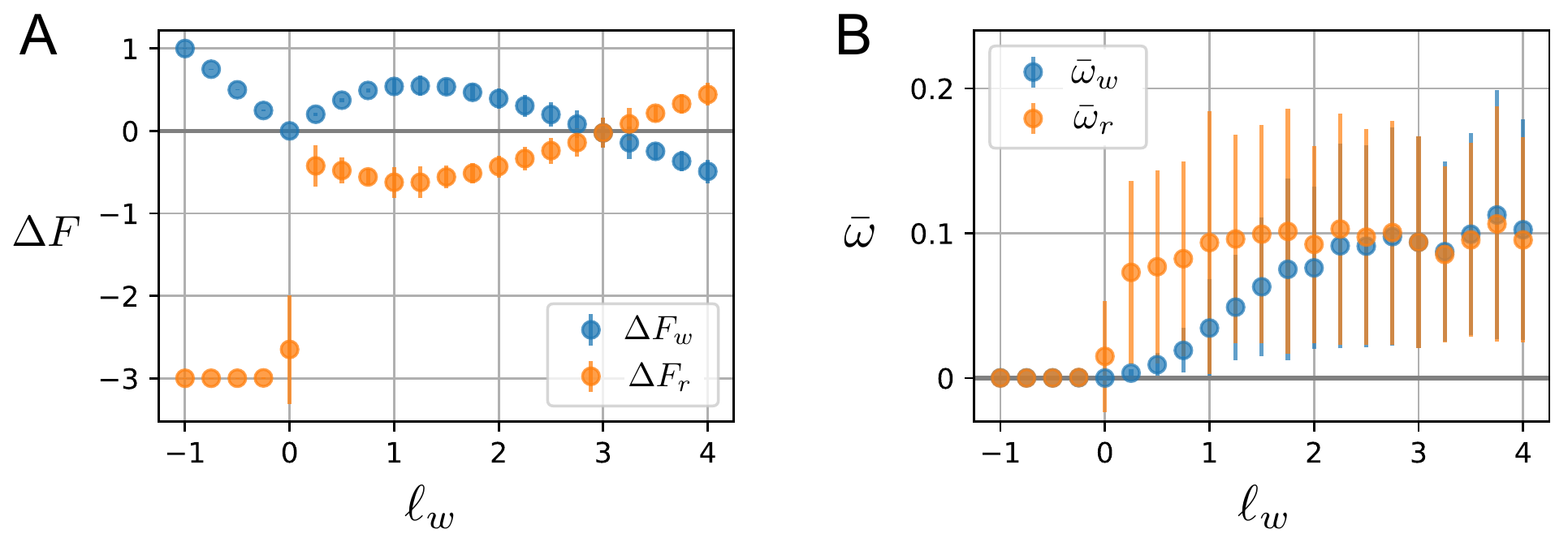}
\caption{Dependence on the ligand $\ell_w$. {\bf A.} Binding free energies $\Delta F_r$ and $\Delta F_w$ as a function of $\ell_w$ for $\ell_0=0$ and $\ell_r=3$ (Fig.~\ref{fig:Nheff}-\ref{fig:ex} correspond to $\ell_w=2$). For $\ell_w<\ell_0$, evolved systems verify $\phi=\min(-\Delta F_r ,\Delta F_w)=\Delta F_w<-\Delta F_r$, while for $\ell_w>\ell_0$, $\phi=\min(-\Delta F_r ,\Delta F_w)=-\Delta F_r=\Delta F_w$ and the two constraints are equally at play. When $\ell_w>\ell_r$, achieving $\phi>0$ is not possible. {\bf B.} Extensions of the conformational switches $\bar\omega_r$ and $\bar\omega_w$ as function of $\ell_w$ for the same systems. For $\ell_w<\ell_0$, no switch occurs ($\bar\omega_r=\bar\omega_w=0$), while for $\ell_w>\ell_0$ switches are obtained that display a wide range of extensions. For each value of $\ell_w$, the dots represent a mean and the error bars a standard deviation over 100 different evolved systems (Fig.~\ref{fig:Nheff}A shows the full distribution of $\bar\omega_r$ for $\ell_w=2$).\label{fig:var_hp}}
\end{center} 
\end{figure}

\begin{figure}[t]
\begin{center}
\includegraphics[width=.7\linewidth]{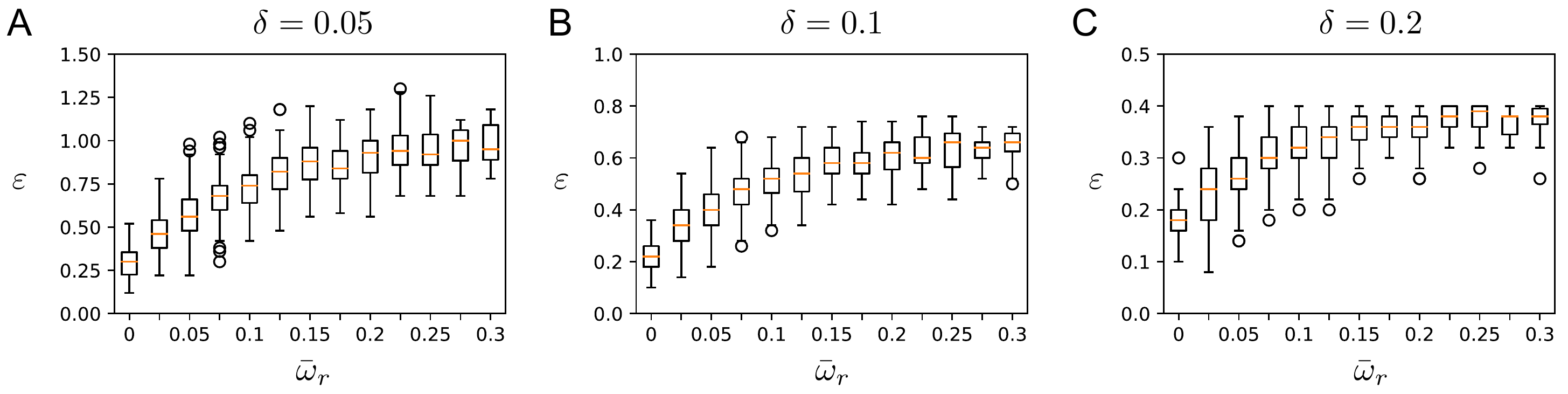}
\caption{{\col Graphs of Fig.~\ref{fig:Nheff}B for alternative choices of the interval $\delta$ used to resolve the different phenotypes. The scale (y-axis) differs in each case but the same trend is observed irrespectively of $\delta$: systems with larger $\bar\omega_r$ have significantly larger $\epsilon$.} \label{FigS:delta}}
\end{center} 
\end{figure}

\begin{figure}[t]
\begin{center}
\includegraphics[width=.5\linewidth]{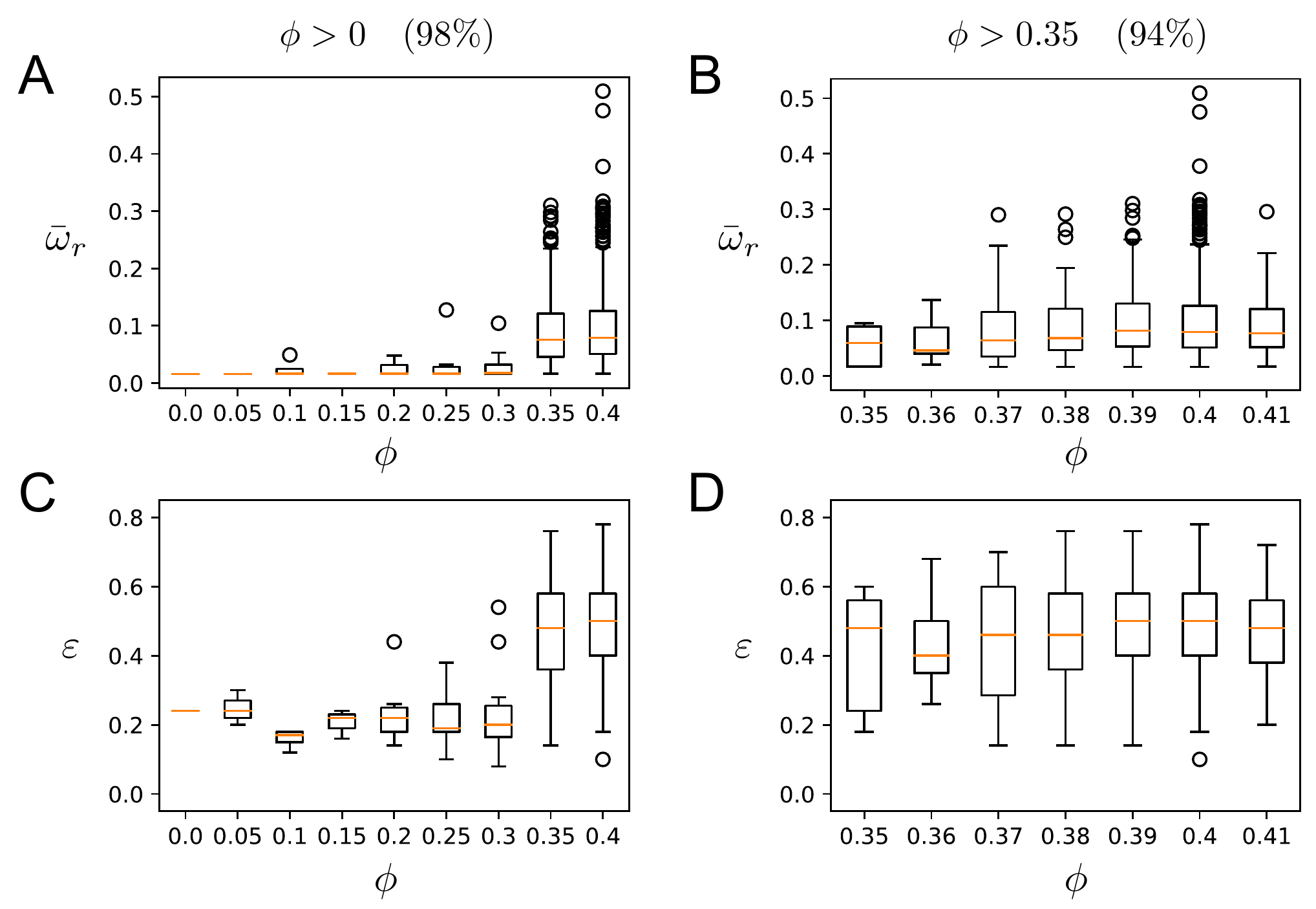}
\caption{Relationships between switch extension, evolvability and fitness -- {\bf A-B.} Switch extension $\bar\omega_r(a)$ as a function of the fitness $\phi(a)$ over $1000$ different evolved systems, showing either the systems with $\phi(a)>0$ (98\% of the total, panel A) or those with $\phi(a)>0.35$ (94\% of the total, panel B). For these fittest systems, on which Fig.~\ref{fig:Nheff} is based, $\bar\omega_r(a)$ is nearly independent of $\phi(a)$. {\bf C-D.} Evolvability $\varepsilon(a)$ as a function of the fitness $\phi(a)$ for the same systems.  For the fittest systems ($\phi(a)>0.35$, panel D), $\varepsilon(a)$ is also nearly independent of $\phi(a)$.\label{FigS:w_phi}}
\end{center} 
\end{figure}

\begin{figure}[t]
\begin{center}
\includegraphics[width=.75\linewidth]{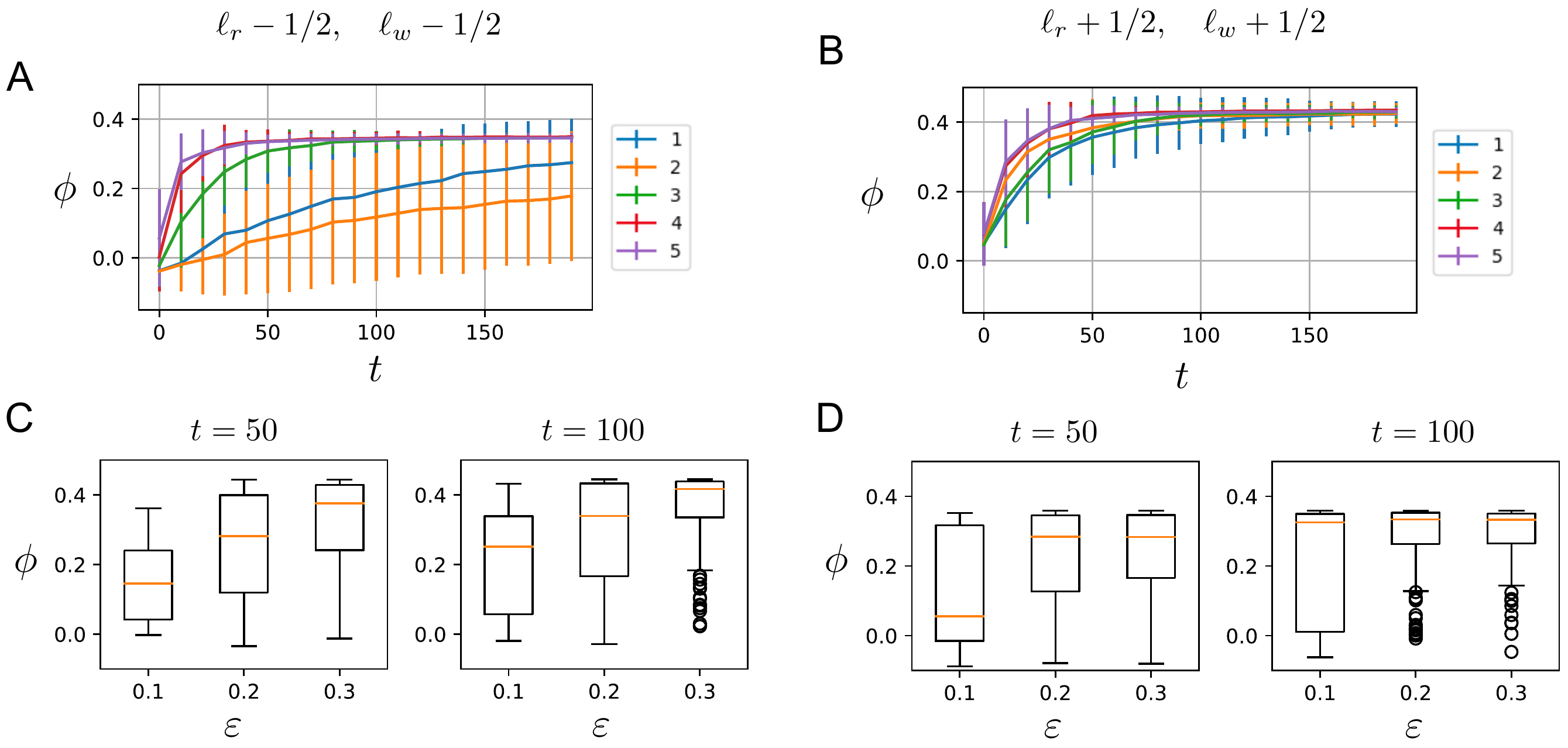}
\caption{{\cl Evolutionary response to a change of selective pressures of systems with different sector sizes and evolvability $\varepsilon$ -- {\bf A.}~Evolutionary trajectories consecutive to a change of selective pressure at $t=0$ from $(\ell_r,\ell_w)=(3,2)$ to $(\ell_r,\ell_w)=(2.5,1.5)$, starting from the five systems of Fig.~\ref{fig:ex}, here labelled 1 to 5. The graph shows the average fitness $\phi$ and its standard deviation over 100 independent trajectories as a function of the number $t$ of generations. It illustrates how the systems with a large sector (3,4,5) adapt more efficiently than the systems with a small sector (1,2). {\bf B.} Same as in A but for a change of selective pressure from $(\ell_r,\ell_w)=(3,2)$ to $(\ell_r,\ell_w)=(3.5,2.5)$. {\bf C.} More systematic analysis starting from the $>940$ systems with $\phi>0.35$ considered in Fig.~S\ref{FigS:w_phi}B,D and performing at $t=0$ the same change of selective pressure as in A. The fitnesses of the systems after $t=50$ and $t=100$ generations are compared to their evolvability $\varepsilon$. This shows that $\varepsilon$, despite being defined based on the effect of single mutations only, is informative of the capacity of a system to adapt over multiple generations. {\bf D.} Same as in C for the change of selective pressure used in B.}\label{FigS:change}}
\end{center} 
\end{figure}

\begin{figure}[t]
\begin{center}
\includegraphics[width=.6\linewidth]{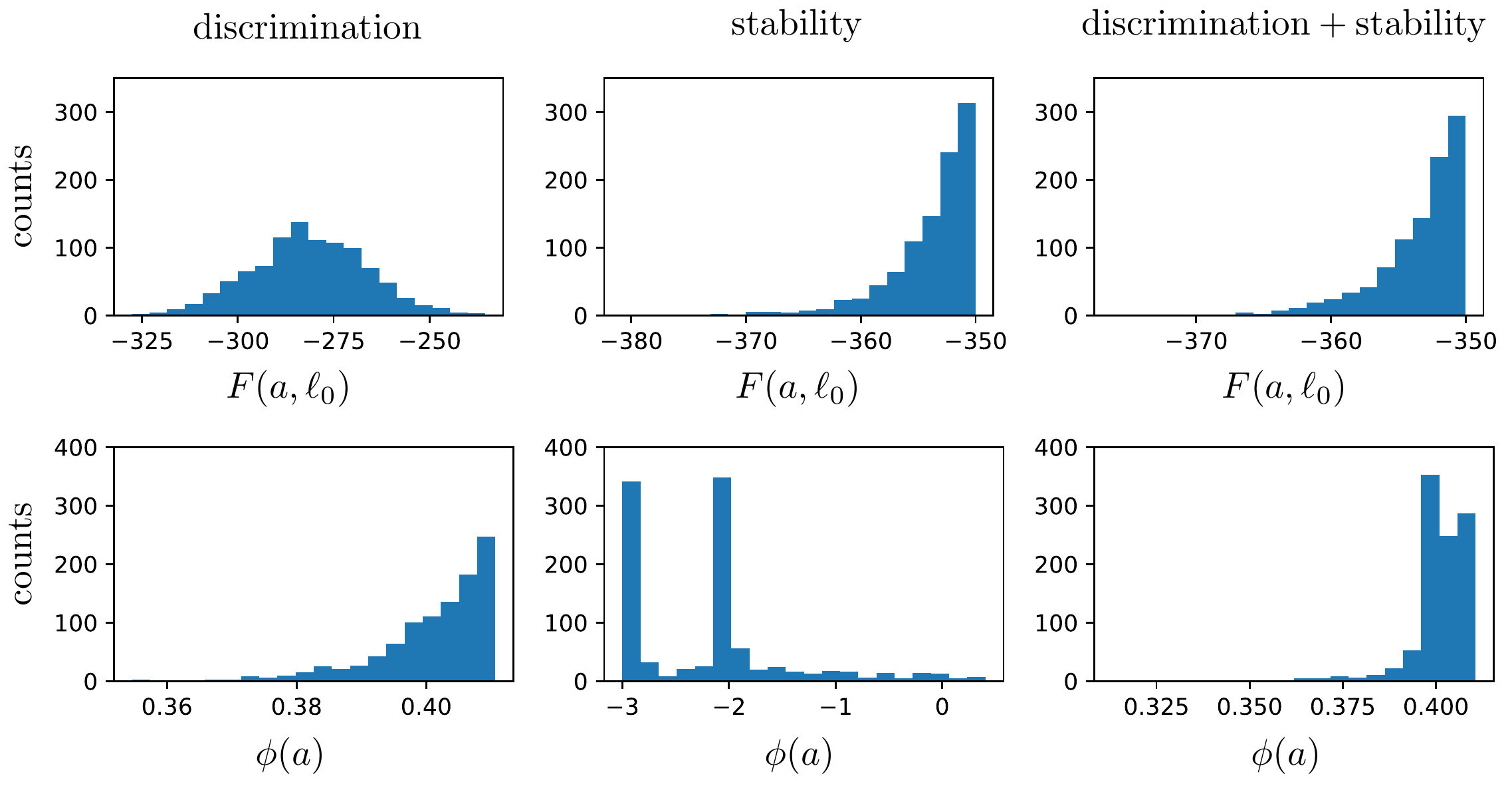}
\caption{Distribution of the free energy $F(a,\ell_0)$ and fitness $\phi(a)$ over the sequences of the three alignments studied in Fig.~\ref{fig:alg} -- The first column corresponds to an alignment selected for exquisite discrimination: the sequences have relatively high free energy, $F(a,\ell_0)>F^*=-350$ and nearly optimal fitness $\phi(a)\simeq 0.41$. The second column corresponds to an alignment generated under the constraint for stability $F(a,\ell_0)<F^*=-350$: the sequences have a free energy $F(a,\ell_0)$ close to the stability threshold $F^*$ but low fitness $\phi(a)<0$. The third column corresponds to an alignment selected for exquisite discrimination under a constraint for stability: the sequences have both a free energy below $F^*$ and a near optimal fitness. \label{FigS:stab_phi}}
\end{center} 
\end{figure}

\begin{figure}[t]
\begin{center}
\includegraphics[width=.25\linewidth]{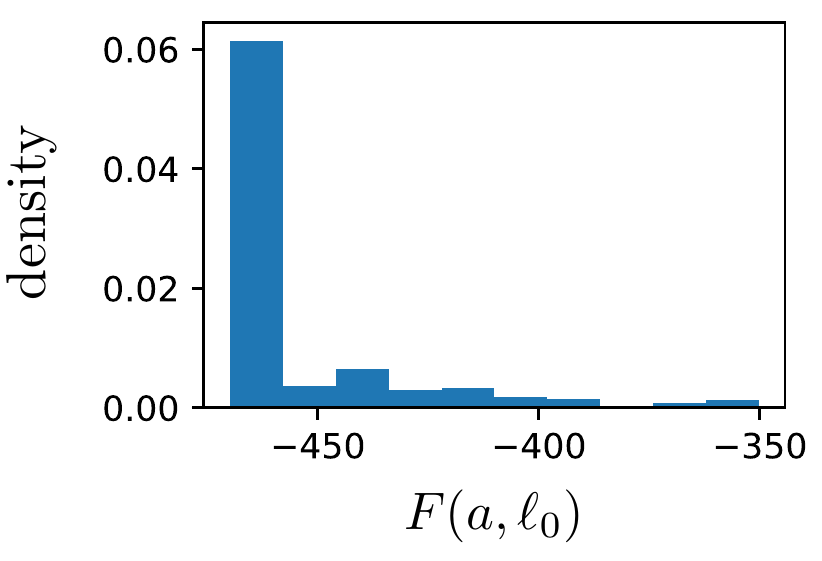}
\caption{Free energy of systems optmized for stability, i.e., evolved with $-F(a,\ell_0)$ as fitness function rather than $\phi(a)$ -- The resulting free energies reach $F_{\rm min}\simeq-450$, far below the free energies of system selected for discrimination without stability constraint ($F(a,\ell_0)>F_0=-325$ in Fig.~\ref{FigS:stab_phi}). To obtain marginally stable systems, the stability threshold $F^*$ is chosen to satisfy $F_{\rm min}<F^*<F_0$. In Fig.~\ref{fig:alg}, $F^*=-350$ with other choices being considered in Fig.~\ref{FigS:Fstar}.\label{FigS:stab_opt}}
\end{center} 
\end{figure}

\begin{figure}[t]
\begin{center}
\includegraphics[width=.6\linewidth]{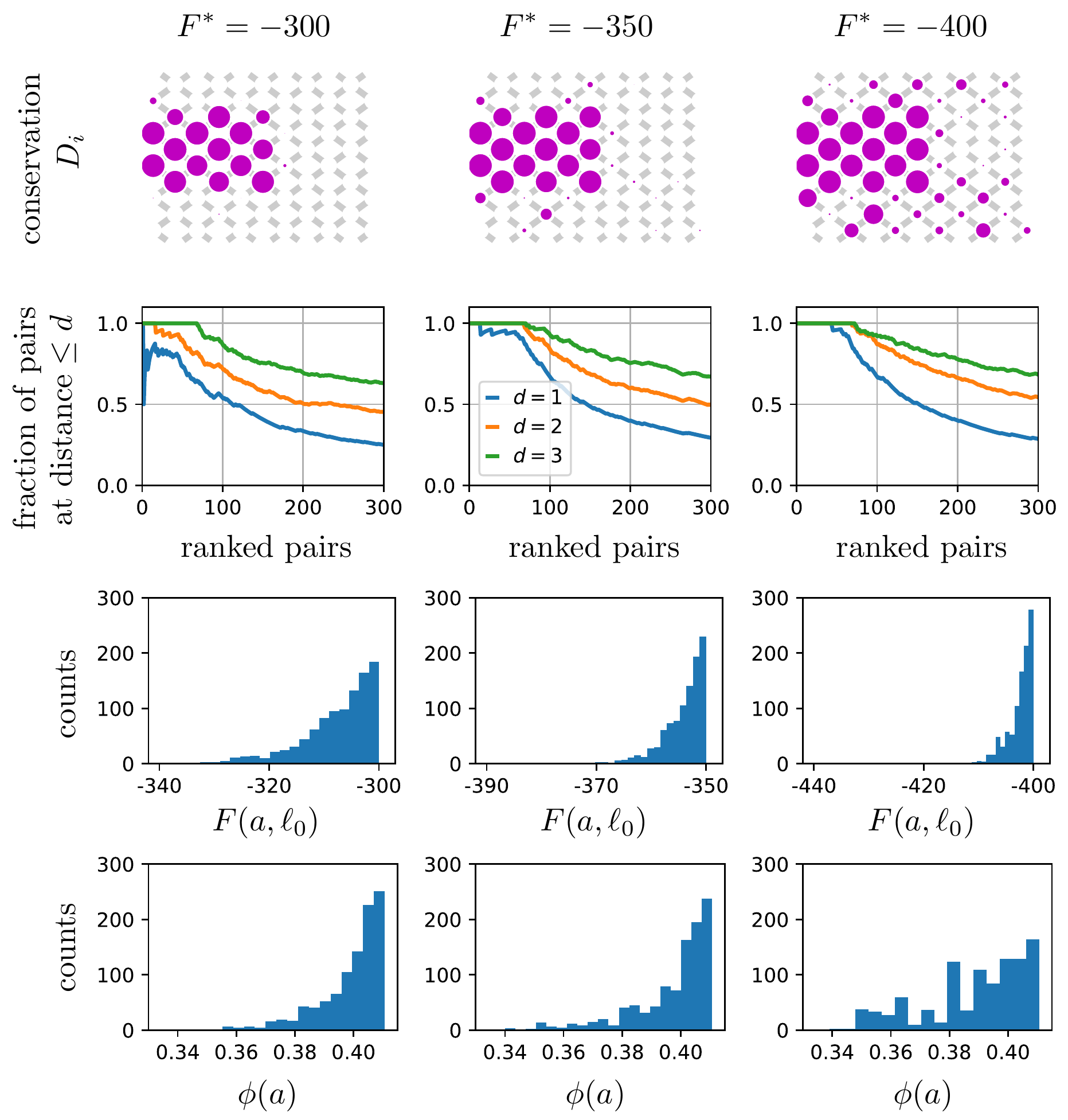}
\caption{Varying the stability threshold $F^*$ -- Similar to Fig.~\ref{fig:alg} and Fig.~\ref{FigS:stab_phi} but considering only alignments produced under both selection for exquisite discrimination and constraint for stability $F(a,\ell_0)<F^*$ ("discimination+stability"). The three alignments are generated from a common sequence with $F(a,\ell_0)<-400$ that is different from the ancestral sequence of alignments in Fig.~\ref{fig:alg}, and then evolved using different values of $F^*$, as indicated on the top. In each case, a sector appears as evolutionary conserved. Coevolving contacts are inferred in every case, although in larger number when the constraint for stability is higher ($F^*$ smaller). \label{FigS:Fstar}}
\end{center} 
\end{figure}

\begin{figure}[t]
\begin{center}
\includegraphics[width=.7\linewidth]{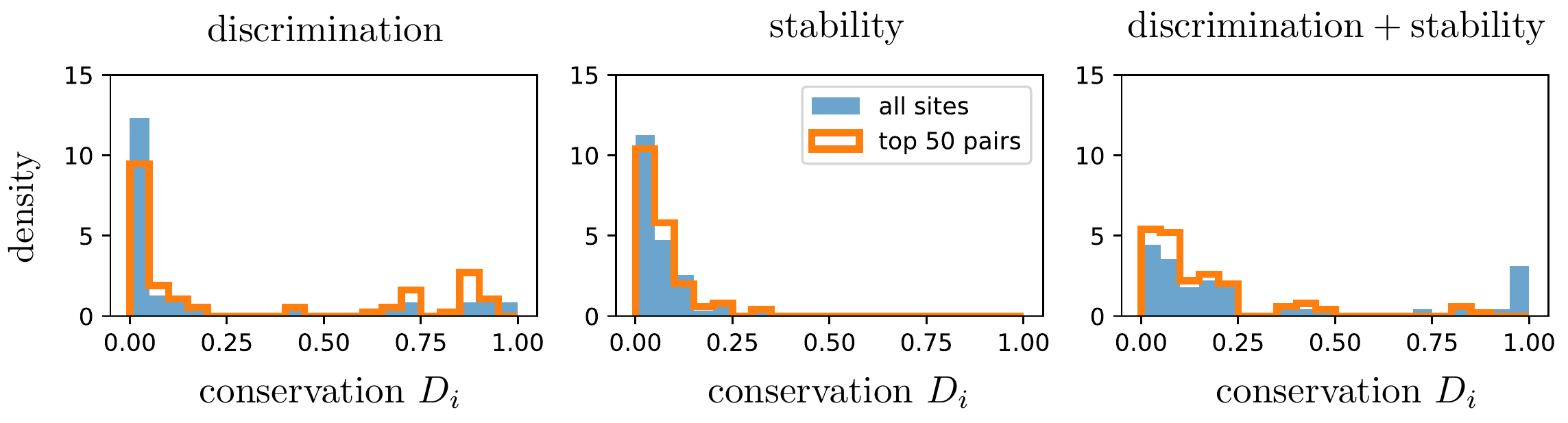}
\caption{Evolutionary conservation of the top 50 pairs returned by DCA for the three alignments of Fig.~\ref{fig:alg}. In blue, distribution of the conservation $D_i$ %given by \eqref{eq:Di} 
over all sites. In orange, distribution of $D_i$ over the 100 sites involved in the top 50 DCA pairs (sites involved in multiple pairs are counted multiple times). Except for the extremely conserved sites of the alignment selected for discrimination and stability (last graph), the two distributions are very similar. This indicates that the coevolving pairs predicted by DCA are distributed all over the structure, independently of the sector inferred by SCA that corresponds to the most conserved sites (Fig.~\ref{fig:ex}).\label{FigS:pairs_D}}
\end{center} 
\end{figure}

\begin{figure}[t]
\begin{center}
\includegraphics[width=.6\linewidth]{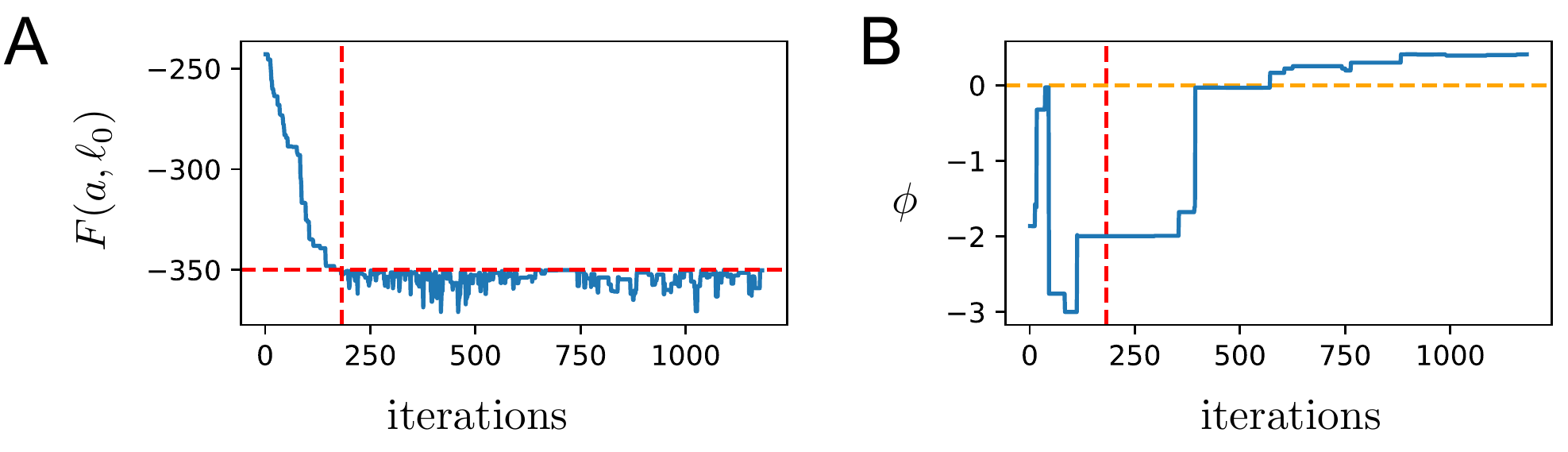}
\caption{Evolution of a stable and discriminating system -- First, the system is evolved by Metropolis Monte Carlo with selection for thermal stability, using $-F(a,\ell_0)$ as fitness function. The iterations are stopped as soon as the required minimal stability, here $F(a,\ell_0)<F^*=-350$, is reached (red dotted lines). $T=1000$ additional iterations are then performed with selection for discrimination and constraint on stability, using $\phi(a)$ as fitness function and excluding any mutation leading to $F(a,\ell_0)>F^*$. {\bf A.} Evolution of $F(a,\ell_0)$. {\bf B.} Evolution of $\phi(a)$. The end-point of this evolution is the system characterized in the second column of Fig.~\ref{fig:ex}, which serves as the common ancestral sequence for  the three alignments of Fig.~\ref{fig:alg}.\label{FigS:time_stab}}
\end{center} 
\end{figure}

\begin{figure}[t]
\begin{center}
\includegraphics[width=.8\linewidth]{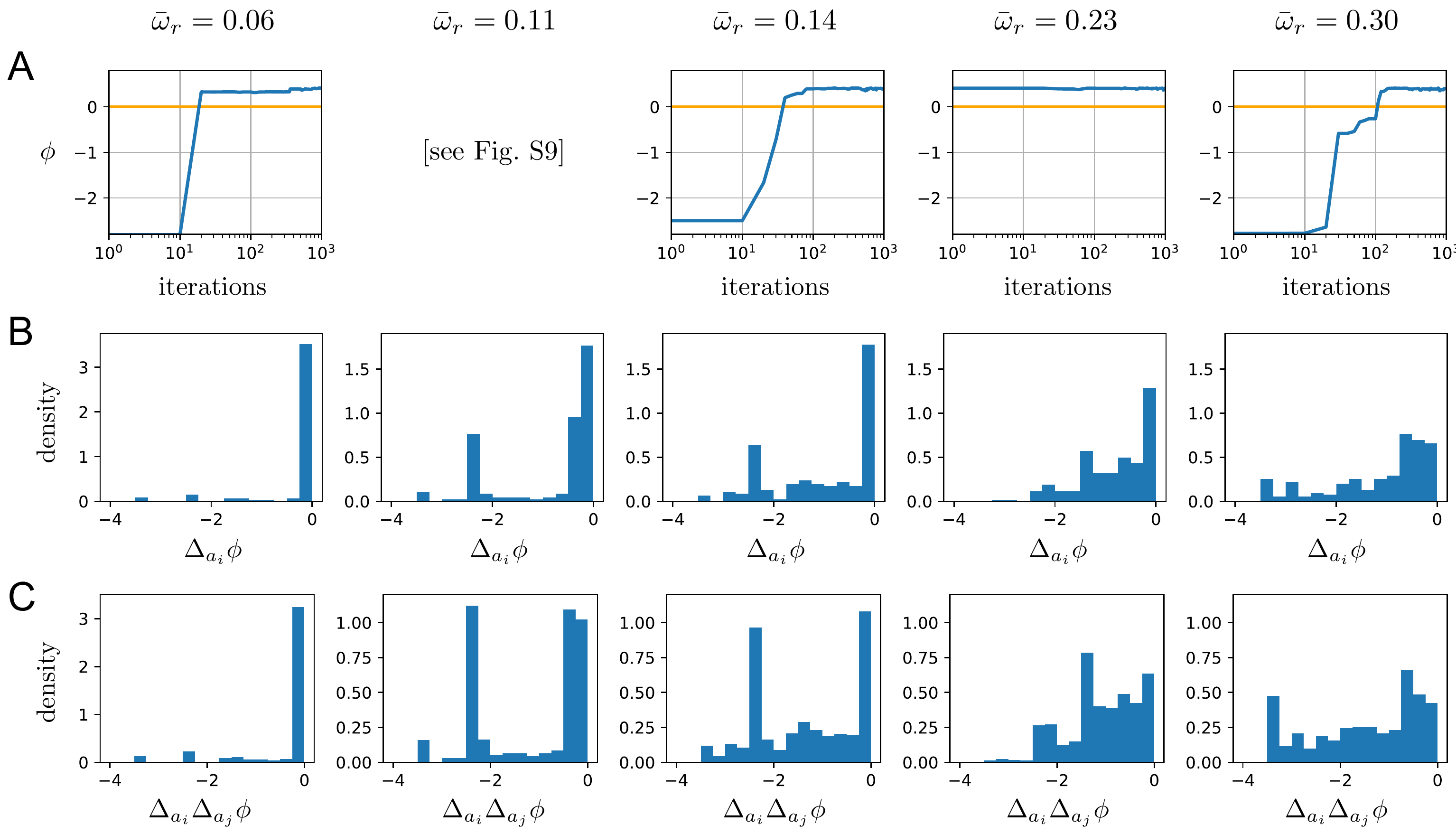}
\caption{Further characterization of the 5 systems presented in Fig.~\ref{fig:ex} -- {\bf A.} Evolution of the fitness $\phi(a)$ as a function of the number of iterations in the Metropolis Monte Carlo algorithm. The systems presented in Fig.~\ref{fig:ex} are the end-point of these trajectories ($T=1000$). The initial conditions are random sequences, except for the second system ($\bar\omega_r=0.11$), which, for the purpose of Fig.~\ref{fig:alg}, is started from a stable sequence satisfying $F(a,\ell_0)<F^*=-350$ (Fig.~\ref{FigS:time_stab}). All systems converge to $\phi\simeq 0.4$, near the theoretical maximum $\max_a\phi(a)=0.41$, although with different extensions $\bar\omega_r$ of their sector. {\bf B.} Distributions of mutational effects $\Delta_{a_i}\phi$ when considering all $(q-1)W(L+1)=275$ possible single mutations. In Fig.~\ref{fig:ex}, the sizes of the blue dots in the last row indicate for each $i$ the mean value of $\Delta_{a_i}\phi$ over the $q-1$ possible mutations. The distributions share a  peak around 0 (neutral mutations) and a more or less extended negative tail, commensurate with $\bar\omega_r$. The peak around  $\Delta_{a_i}\phi\simeq -2.4$ that is observed in some cases corresponds to mutants that are effectively equivalent to a single-spin model with $a'=0$, in which case $\phi(a')=\min(|\ell_r|,-|\ell_w|)=-\ell_w=-2$.
\label{FigS:time}}
\end{center} 
\end{figure}

\end{widetext}


\begin{thebibliography}{10}

\bibitem{licata1995long}
LiCata VJ, Ackers GK (1995) Long-range, small magnitude nonadditivity of
  mutational effects in proteins.
\newblock {\em Biochemistry} 34(10):3133--3139.

\bibitem{Daugherty:2000gz}
Daugherty PS, Chen G, Iverson BL, Georgiou G (2000) {Quantitative analysis of
  the effect of the mutation frequency on the affinity maturation of single
  chain Fv antibodies}.
\newblock {\em Proceedings of the National Academy of Sciences}
  97(5):2029--2034.

\bibitem{Morley:2005kc}
Morley KL, Kazlauskas RJ (2005) {Improving enzyme properties: when are closer
  mutations better?}
\newblock {\em Trends in Biotechnology} 23(5):231--237.

\bibitem{Lockless:1999uf}
Lockless SW, Ranganathan R (1999) {Evolutionarily conserved pathways of
  energetic connectivity in protein families.}
\newblock {\em Science} 286(5438):295--299.

\bibitem{Halabi:2009jc}
Halabi N, Rivoire O, Leibler S, Ranganathan R (2009) {Protein sectors:
  evolutionary units of three-dimensional structure.}
\newblock {\em Cell} 138(4):774--786.

\bibitem{Morcos:2011jg}
Morcos F, et~al. (2011) {Direct-coupling analysis of residue coevolution
  captures native contacts across many protein families.}
\newblock {\em Proceedings of the National Academy of Sciences} 108(49):E1293--301.

\bibitem{fersht1985enzyme}
Fersht A (1985) {\em Enzyme structure and mechanism}.
\newblock (WH Freeman) Vol.{}~99.

\bibitem{pawson2000protein}
Pawson T, Nash P (2000) Protein--protein interactions define specificity in
  signal transduction.
\newblock {\em Genes \& development} 14(9):1027--1047.

\bibitem{todeschini2014transcription}
Todeschini AL, Georges A, Veitia RA (2014) Transcription factors: specific DNA
  binding and specific gene regulation.
\newblock {\em Trends in genetics} 30(6):211--219.

\bibitem{VanRegenmortel:1998uv}
Van~Regenmortel MH (1998) {From absolute to exquisite specificity. Reflections
  on the fuzzy nature of species, specificity and antigenic sites.}
\newblock {\em Journal of immunological methods} 216(1-2):37--48.

\bibitem{dubay2011long}
DuBay KH, Bothma JP, Geissler PL
{Long-range intra-protein communication can be transmitted by correlated side-chain fluctuations alone}.
\newblock {\em PLoS Comp Biol} 7(9): e1002168.

\bibitem{spandrel}
Gould SJ, Lewontin RC (1979) {The spandrels of San Marco and the Panglossian paradigm: a critique of the adaptationist programme.}
\newblock {\em Proc Royal Soc B} 205 (1161):58--598.

\bibitem{Sanejouand:2013kj}
Sanejouand YH (2013) {Elastic network models: theoretical and empirical
  foundations}.
\newblock {\em Methods in molecular biology } 924 (Chapter
  23):601--616.

\bibitem{Zadorin19}
Zadorin AS (2019)
{Allostery and conformational changes upon binding as generic features of proteins: a high-dimension geometrical approach.} \newblock {\em arXiv}:1905.02815.

\bibitem{Hemery:2015ei}
Hemery M, Rivoire O (2015) {Evolution of sparsity and modularity in a model of
  protein allostery.}
\newblock {\em Physical review E} 91(4):042704--10.

\bibitem{metropolis53}
Metropolis N, Rosenbluth AW, Rosenbluth MN, Teller AH, Teller E (1953) Equation
  of state calculations by fast computing machines.
\newblock {\em The journal of chemical physics} 21(6):1087--1092.

\bibitem{McCandlish:2014us}
McCandlish DM, Stoltzfus A (2014) {Modeling evolution using the probability of
  fixation: history and implications}.
\newblock {\em The Quarterly Review of Biology} 89(3):225--252.

\bibitem{baxter82}
Baxter RJ (1982) {\em Exactly solved models in statistical mechanics}.
\newblock (Academic Press).

\bibitem{Rivoire:2013cy}
Rivoire O (2013) {Elements of coevolution in biological sequences.}
\newblock {\em Physical Review Letters} 110(17):178102--5.

\bibitem{Cocco:2013el}
Cocco S, Monasson R, Weigt M (2013) {From Principal Component to Direct
  Coupling Analysis of Coevolution in Proteins: Low-Eigenvalue Modes are Needed
  for Structure Prediction}.
\newblock {\em PLoS computational biology} 9(8):e1003176.

\bibitem{Rivoire:2016bl}
Rivoire O, Reynolds KA, Ranganathan R (2016) {Evolution-Based Functional
  Decomposition of Proteins}.
\newblock {\em PLoS computational biology} 12(6):e1004817.

\bibitem{Taverna:2002gj}
Taverna DM, Goldstein RA (2002) {Why are proteins marginally stable?}
\newblock {\em Proteins: Structure, Function, and Bioinformatics}
  46(1):105--109.

\bibitem{Ekeberg:2013fq}
Ekeberg M, L{\"o}vkvist C, Lan Y, Weigt M, Aurell E (2013) {Improved contact
  prediction in proteins: using pseudolikelihoods to infer Potts models.}
\newblock {\em Physical review E} 87(1):012707.

\bibitem{flechsig2017design}
Flechsig H (2017) Design of elastic networks with evolutionary optimized
  long-range communication as mechanical models of allosteric proteins.
\newblock {\em Biophysical journal} 113(3):558--571.

\bibitem{rocks2017designing}
Rocks JW, Pashine N, Bischofberger I, Goodrich CP, Liu AJ, Nagel SR  (2017) Designing allostery-inspired response in mechanical
  networks.
\newblock {\em Proceedings of the National Academy of Sciences}
  114(10):2520--2525.

\bibitem{tlusty2017physical}
Tlusty T, Libchaber A, Eckmann JP (2017) Physical model of the
  genotype-to-phenotype map of proteins.
\newblock {\em Physical Review X} 7(2):021037.

\bibitem{yan2017architecture}
Yan L, Ravasio R, Brito C, Wyart M (2017) Architecture and coevolution of
  allosteric materials.
\newblock {\em Proceedings of the National Academy of Sciences}
  114(10):2526--2531.

\bibitem{Kuriyan:2007}
Kuriyan J, Eisenberg D (2007)
{The origin of protein interactions and allostery in colocalization.}
\newblock {\em Nature} 450(7172), 983.

\bibitem{Gunasekaran:2004}
Gunasekaran K, Ma B, Nussinov R (2004)
{Is allostery an intrinsic property of all dynamic proteins?}. 
\newblock {\em Proteins: Structure, Function, and Bioinformatics} 57(3), 433--443.

\bibitem{Hardy:2004gy}
Hardy JA, Wells JA (2004) {Searching for new allosteric sites in enzymes.}
\newblock {\em Current Opinion in Structural Biology} 14(6):706--715.

\bibitem{zorn:2010}
Zorn JA, Wells JA (2010). 
{Turning enzymes ON with small molecules.}
\newblock {\em Nature chemical biology} 6(3), 179.

\bibitem{Lee:2008gd}
Lee J, et~al. (2008) {Surface sites for engineering allosteric control in
  proteins.}
\newblock {\em Science} 322(5900):438--442.

\bibitem{Reynolds:2011gs}
Reynolds KA, McLaughlin RN, Ranganathan R (2011) {Hot Spots for Allosteric
  Regulation on Protein Surfaces}.
\newblock {\em Cell} 147(7):1564--1575.

\bibitem{pincus2017evolution}
Pincus D, Pandey JP, Feder ZA, Creixell P, Resnekov O, Reynolds KA (2018). 
Engineering allosteric regulation in protein kinases. 
\newblock {\em Sci. Signal.} 11(555):eaar3250.

\bibitem{Coyle:2013dn}
Coyle SM, Flores J, Lim WA (2013) {Exploitation of latent allostery enables the
  evolution of new modes of MAP kinase regulation.}
\newblock {\em Cell} 154(4):875--887.

\bibitem{Ancel:2000gj}
Ancel LW, Fontana W (2000) {Plasticity, evolvability, and modularity in RNA.}
\newblock {\em The Journal of experimental zoology} 288(3):242--283.

\bibitem{Parter:2008bb}
Parter M, Kashtan N, Alon U (2008) {Facilitated Variation: How Evolution Learns
  from Past Environments To Generalize to New Environments}.
\newblock {\em PLoS computational biology} 4(11):e1000206.

\bibitem{Raman:2016gv}
Raman AS, White KI, Ranganathan R (2016) {Origins of Allostery and Evolvability
  in Proteins: A Case Study.}
\newblock {\em Cell} 166(2):468--480.
{\cl
\bibitem{savir2007conformational}
Savir Y, Tlusty T (2007)
{Conformational proofreading: the impact of conformational changes on the specificity of molecular recognition.}
\newblock {\em PloS one} 2(5):e468.

\bibitem{vlaminck}
De Vlaminck I, van Loenhout MT, Zweifel L, den Blanken J, Hooning K, Hage S, Kerssemakers J, Dekker C. (2012)
{Mechanism of homology recognition in DNA recombination from dual-molecule experiments.}
\newblock {\em Molecular cell} 46(5):616-624.
}
\bibitem{Miller:1997es}
Miller DW, Dill KA (1997) {Ligand binding to proteins: the binding landscape
  model}.
\newblock {\em Protein science}
  6(10):2166--2179.

\bibitem{Williams:2001ua}
Williams PD, Pollock DD, Goldstein RA (2001) {Evolution of functionality in
  lattice proteins}.
\newblock {\em Journal of molecular graphics {\&} modelling} 19(1):150--156.

\bibitem{Manhart:2015eg}
Manhart M, Morozov AV (2015) {Protein folding and binding can emerge as
  evolutionary spandrels through structural coupling.}
\newblock {\em Proceedings of the National Academy of Sciences} 112(6):1797--1802.
{\cl
\bibitem{Morcos14}
Morcos F, Schafer NP, Cheng RR, Onuchic JN, Wolynes PG (2014)
{Coevolutionary information, protein folding landscapes, and the thermodynamics of natural selection.} 
\newblock {\em Proceedings of the National Academy of Sciences} 111(34):12408--2413.

\bibitem{Figliuzzi15}
Figliuzzi M, Jacquier H, Schug A, Tenaillon O, Weigt M (2015). 
{Coevolutionary landscape inference and the context-dependence of mutations in beta-lactamase TEM-1.}
\newblock {\em Molecular biology and evolution} 33(1):268--280.
}
\bibitem{Jacquin:2016cl}
Jacquin H, Gilson A, Shakhnovich E, Cocco S, Monasson R (2016) {Benchmarking
  Inverse Statistical Approaches for Protein Structure and Design with Exactly
  Solvable Models.}
\newblock {\em PLoS computational biology} 12(5):e1004889.

\bibitem{McLaughlinJr:2012hw}
McLaughlin~Jr RN, Poelwijk FJ, Raman A, Gosal WS, Ranganathan R (2012) {The
  spatial architecture of protein function and adaptation}.
\newblock {\em Nature} 491(7422):138--142.

\bibitem{Motlagh:2014kc}
Motlagh HN, Wrabl JO, Li J, Hilser VJ (2014) {The ensemble nature of allostery}.
\newblock {\em Nature} 508 (7496):331--339.

\bibitem{Saavedra:2018ea}
Saavedra HG, Wrabl JO, Anderson JA, Li J, Hilser VJ (2018) {Dynamic allostery can drive cold adaptation in enzymes}.
\newblock {\em Nature} 558 (7709):324--328.

\bibitem{Cooper:1984cn}
Cooper A, Dryden DTF (1984) {Allostery without conformational change.}
\newblock {\em European Biophysics Journal} 11(2):103--109.

\bibitem{Popovych:2006}
Popovych N, Sun S, Ebright RH, Kalodimos CG (2006). {Dynamically driven protein allostery.}
\newblock {\em Nature structural \& molecular biology} 13(9): 831.

\bibitem{mcleish:2013}
McLeish TCB, Rodgers TL, Wilson MR (2013) {Allostery without conformation change: modelling protein dynamics at multiple scales}.
\newblock {\em Phys Biol} 10(5):056004.

\bibitem{Townsend:2015}
Townsend PD {\it et al.} (2015) {The role of protein-ligand contacts in allosteric regulation of the Escherichia coli catabolite activator protein.}
\newblock {\em Journal of Biological Chemistry} 290(36): 22225--22235.

\bibitem{Socolich:2005js}
Socolich M, et~al. (2005) {Evolutionary information for specifying a protein
  fold}.
\newblock {\em Nature} 437(7058):512--518.

\bibitem{Russ:2005kc}
Russ WP, Lowery DM, Mishra P, Yaffe MB, Ranganathan R (2005) {Natural-like
  function in artificial WW domains}.
\newblock {\em Nature} 437(7058):579--583.

\bibitem{figliuzzi2018pairwise}
Figliuzzi M, Barrat-Charlaix P, Weigt M (2018) How pairwise coevolutionary
  models capture the collective residue variability in proteins?
\newblock {\em Molecular biology and evolution} 35(4):1018--1027.

\bibitem{Smith:1997vn}
Smith GP, Petrenko VA (1997) {Phage Display}.
\newblock {\em Chemical Reviews} 97(2):391--410.

\bibitem{Levin:2006}
Levin AM, Weiss GA (2006). 
{Optimizing the affinity and specificity of proteins with molecular display.}
 \newblock {\em Molecular Biosystems} 2(1):49--57.

\bibitem{Nizak:2003}
Nizak C, Monier S, del Nery E, Moutel S, Goud B, Perez F (2003). 
{Recombinant antibodies to the small GTPase Rab6 as conformation sensors.}
\newblock {\em  Science} 300(5621), 984--987.

\bibitem{Garcia:2010}
Garcia-Rodriguez C et al. 2010).
{Neutralizing human monoclonal antibodies binding multiple serotypes of botulinum neurotoxin.}
\newblock {\em  Protein engineering, design \& selection} 24(3): 321--331.

\bibitem{Fowler:2014gq}
Fowler DM, Fields S (2014) {Deep mutational scanning: a new style of protein
  science.}
\newblock {\em Nature Methods} 11(8):801--807.


\end{thebibliography}

\begin{thebibliography}{10}

\bibitem{Perelson:1979gb}
Perelson AS, Oster GF (1979) {Theoretical studies of clonal selection: minimal
  antibody repertoire size and reliability of self-non-self discrimination.}
\newblock {\em Journal of theoretical biology} 81(4):645--670.

\bibitem{Ekeberg:2013fq_re}
Ekeberg M, L{\"o}vkvist C, Lan Y, Weigt M, Aurell E (2013) {Improved contact
  prediction in proteins: using pseudolikelihoods to infer Potts models.}
\newblock {\em Physical review E} 87(1):012707.

\bibitem{juliaDCA}
Pagnani A (2018) {Pseudo Likelihood Maximization for protein in Julia. \url{https://github.com/pagnani/PlmDCA}}

\bibitem{Rivoire:2016blre}
Rivoire O, Reynolds KA, Ranganathan R (2016) {Evolution-Based Functional
  Decomposition of Proteins}.
\newblock {\em PLoS computational biology} 12(6):e1004817.

\end{thebibliography}
\end{document}